\documentclass[11pt]{article}
\pdfoutput=1
\usepackage{jheppub,amsmath,amssymb}
\usepackage{enumerate}
\usepackage{bm}
\usepackage{bbm}
\usepackage[usenames,dvipsnames]{xcolor}
\usepackage{graphics}
\usepackage{tikz,todonotes}
\usepackage[utf8]{inputenc}
\usetikzlibrary{arrows,positioning,decorations.markings,decorations.pathmorphing,calc}

\definecolor{hyperref}{RGB}{026,028,087}
\newcommand{\p}{{\bf p}}

\def\gsim{ \lower .75ex \hbox{$\sim$} \llap{\raise .27ex \hbox{$>$}} }
\def\lsim{ \lower .75ex \hbox{$\sim$} \llap{\raise .27ex \hbox{$<$}} }
\def\be{\begin{equation}}
\def\ee{\end{equation}}
\def\bea{\begin{eqnarray}}
\def\eea{\end{eqnarray}}
\newcommand{\nn}{\nonumber}

\def \bal#1\eal  {\begin{align} #1 \end{align}}
\newcommand{\ud} {\mathrm{d}}

\newcommand{\pd} {\partial}

\newcommand{\ai}{{\alpha}}
\newcommand{\bi}{{\beta}}

\newcommand{\li}{{\lambda}}

\newcommand{\ba}{\begin{array}}
\newcommand{\ea}{\end{array}}
\newcommand{\mn}{\mu\nu}

\newcommand{\commentout}[1]{}

\newcommand{\comment}[1]{}

\newcommand{\bs}{\begin{split}}

\def\ba{\begin{eqnarray}}
\def\ea{\end{eqnarray}}

\def\nn{\nonumber}

\def\d{\mathrm{d}}

\def\({\left(}
\def\){\right)}
\def\ie{{\it i.e. }}

\newcommand*{\mathcolor}{}
\def\mathcolor#1#{\mathcoloraux{#1}}
\newcommand*{\mathcoloraux}[3]{%
  \protect\leavevmode
  \begingroup
    \color#1{#2}#3%
  \endgroup
}
\newlength{\stheight}
\newcommand\textst[1][fu-grey]{
	\ifmmode\setlength{\stheight}{+1.0ex}
	\else\setlength{\stheight}{+0.5ex}
	\fi
	\bgroup\markoverwith{\textcolor{#1}{\rule[\the\stheight]{2pt}{1.0pt}}}\ULon
} 
\newcommand{\textins}[2][fu-grey]{
	\ifmmode\mathcolor{#1}{#2}
	\else\textcolor{#1}{#2}\@\,
	\fi
}

\def\({\left(}
\def\){\right)}

\def\L{{\cal L}}

\allowdisplaybreaks[3]
\def\ba{\begin{eqnarray}}
\def\ea{\end{eqnarray}}

\def\L{\mathcal{L}}

\def\D{\mathcal{D}}

\def\d{\mathrm{d}}
\def\mn{_{\mu \nu}}

\def\({\left(}
\def\){\right)}

\def\p{\partial}
\def\ie{{\em i.e. }}
\def\ien{{\em i.e.}}

\begin{document}

\title{Massive Galileon Positivity Bounds}

%\author{Claudia de Rham$^{a,b}$, Scott Melville$^a$, Andrew J. Tolley$^{a,b}$, Shuang-Yong Zhou$^b$}

\author[a,b]{Claudia de Rham}
\author[a]{Scott Melville}
\author[a,b]{Andrew J. Tolley}
\author[a]{Shuang-Yong Zhou}
\affiliation[a]{Theoretical Physics, Blackett Laboratory, Imperial College, London, SW7 2AZ, U.K.}
\affiliation[b]{CERCA, Department of Physics, Case Western Reserve University, 10900 Euclid Ave, Cleveland, OH 44106, USA}

\emailAdd{c.de-rham@imperial.ac.uk}
\emailAdd{s.melville16@imperial.ac.uk}
\emailAdd{a.tolley@imperial.ac.uk}
\emailAdd{shuangyong.zhou@imperial.ac.uk}

\abstract{
The EFT coefficients in any gapped, scalar, Lorentz invariant field theory must satisfy positivity requirements if there is to exist a local, analytic Wilsonian UV completion. We apply these bounds to the tree level scattering amplitudes for a massive Galileon. The addition of a mass term, which does not spoil the non-renormalization theorem of the Galileon and preserves the Galileon symmetry at loop level, is necessary to satisfy the lowest order positivity bound. We further show that a careful choice of successively higher derivative corrections are necessary to satisfy the higher order positivity bounds. There is then no obstruction to a local UV completion from considerations of tree level 2-to-2 scattering alone. To demonstrate this we give an explicit example of such a UV completion.
}

\maketitle

%\newpage
%\setcounter{tocdepth}{2}
%\tableofcontents

%%%%%%%%%%%%%%%%
\section{Introduction}
%%%%%%%%%%%%%%%%

Low energy effective field theories of scalar fields are part and parcel of cosmological model building. They are a near essential ingredient in inflationary theories, and form the basis of most theories of, or alternatives to, dark energy. In many proposed models, the scalar is an assumed low energy field in an otherwise unknown high energy (UV) completion. In the absence of explicit UV guidance, effective field theories can be constructed according to the standard principle that every operator consistent with the underlying symmetries and field content is included in the Lagrangian. The form of the scalar low-energy effective field theory (LEEFT) is then significantly controlled by the assumed symmetry, be it exact or approximate. A special class of such LEEFT are the Galileon models \cite{Nicolis:2004qq} where the assumed global symmetry for the scalar field $\pi$ is the spacetime dependent transformation $\pi \to \pi + c +v_\mu x^\mu$.
Theories of this type were discovered in the context of massive theories of gravity, originally in the Dvali-Gabadadze-Porrati model \cite{Dvali:2000hr}, where $\pi$ describes the degree of freedom associated with the helicity zero mode of the massive graviton \cite{Luty:2003vm}.  \\

In order for a gapped (i.e. massive) scalar theory to admit a standard Wilsonian UV completion, the 2-to-2 scattering amplitude must satisfy a number of so-called positivity bounds \cite{Adams:2006sv,deRham:2017avq}. These are derived based on the cherished assumptions that the scattering amplitude is Lorentz invariant, unitary, polynomially bounded in momenta, crossing symmetric and is analytic in the complex energy plane modulo certain poles and branch cuts. Of these conditions, the latter two are tied to locality and causality. Polynomial boundedness, the statement that the scattering amplitudes do not grow faster than a given polynomial (or slightly more generally a linear exponential) of complex momenta is necessary so that Fourier transforms are well defined to ensure that the amplitudes can be given meaning in real space. This is tied to locality of real space correlation functions. Analyticity is motivated by causality, and in the special cases where it can be derived rigorously, analyticity follows from the properties of the real space retarded Green's functions which are used to determine the S-matrix amplitude. Although no rigorous proof of full analyticity of the S-matrix has ever been given, it is straightforward to show that to any order in perturbation theory the scattering amplitude remains analytic, and it is generally argued that the singularities (position of poles and branch cuts) on the physical sheet in the full S-matrix are the same as those seen in perturbation theory \cite{Eden:2012}. \\

The lowest order forward limit positivity bounds were previously used to argue that the massless Galileon \cite{Nicolis:2008in} had no standard UV completion, because the coefficient of $s^2$ ($s$ being the square of the center of mass energy) in the (pole subtracted) scattering amplitude which must necessarily be positive definite, was found to be zero \cite{Adams:2006sv}. This unusual behaviour is a consequence of the special soft scattering properties of Galileons, which are in turn tied to the Galileon symmetry $\pi \rightarrow \pi +c +v_{\mu } x^{\mu}$ \cite{Cheung:2014dqa,Cheung:2016drk}. This argument, however, relies on a subtle procedure which introduces a mass as an IR regulator and sends $m\to0$ at the end of the calculation. \\

The standard positivity bounds are only well-defined in the presence of a mass gap. The reason is two-fold: on the one hand a massless theory can violate the Froissart bound \cite{Azimov:2011nk, Diez:2013masters} which affects the number of subtractions necessary. If for example three subtractions were needed, due to the cross section growing faster than $\sigma \sim s$ (something which is technically possible for a massless theory) then it is impossible to place a bound on the sign of the coefficient of the $s^2$ term. Secondly, the mass gap is necessary to have an analytic region for the scattering amplitude which connects the upper and lower half complex $s$ planes. In the massless limit, the amplitude at $t=0$ is ill-defined due to a pole and branch point\footnote{In practice the latter point is less of a concern: the pole can always be subtracted and the branch point at $t=0$ only arises at the loop level and so does not affect bounds on the tree amplitude of the LEEFT.}.   \\

Fortunately these problems are easily dealt with by considering a massive Galileon theory. In its original realization, the Galileon arose as the helicity-zero mode of a massive spin-2 resonance, and thus was neither strictly massless nor massive. It is only in a specific decoupling limit that it becomes massless.  On the other hand, a massive Galileon sector is more naturally embedded in interacting spin-2 theories such as Ghost-free massive gravity \cite{deRham:2009rm,deRham:2010ik,deRham:2010kj}, (see \cite{deRham:2014zqa,deRham:2016nuf} for a recent review of massive gravity)---the scalar mode in this massive gravity theory is massive away from the decoupling limit and corresponds to the helicity-0 component of the massive graviton. Despite na\"{i}vely breaking the Galileon symmetry, a mass term leaves intact the Galileon non-renormalization theorem \cite{Burrage:2011cr}. Furthermore, no additional operators violating the Galileon symmetry are generated at any order in loops by the presence of a mass term. It is therefore very natural to consider a massive Galileon theory, and we will show here that such a theory can indeed satisfy the positive bounds required for an analytic Lorentz invariant UV completion, providing its higher order derivative terms have suitably chosen coefficients. \\

Our results are consistent with the recent findings that forward limit positivity bounds are satisfied for various IR extensions of the Galileon (massive gravity \cite{Cheung:2016yqr}, and pseudo-linear massive gravity / Proca theory \cite{Bonifacio:2016wcb}). While one might conclude from this that the non-scalar modes in Ghost-free massive gravity play an important role in satisfying the positivity requirements, based on the massive Galileon result we see that these requirements are already satisfied for the massive scalar sector alone. In fact, as we are able to apply our positivity bounds away from the forward limit \cite{deRham:2017avq}, our requirements on the massive Galileon are stronger than equivalent bounds in the literature. \\

Starting with the hypothetical existence of a local, analytic Wilsonian UV completion to a Lorentz invariant massive scalar LEEFT, we have derived previously  a series of positivity bounds that the (pole subtracted) $2\to 2$ scattering amplitude and its derivatives with respect to the Mandelstam variables should satisfy, see Ref.~\cite{deRham:2017avq}. These represent an infinite number of requirements that place constraints on all the coefficients of the scattering amplitude when Taylor expanded in terms of the Mandelstam variables.
This Taylor expansion is always well defined given the analyticity of the pole subtracted amplitude in the Mandelstam triangle $0 \le s,t,u \le 4m^2$.
These bounds are valid at and away from the forward scattering limit and as we shall see later, the more stringent bounds are often away from the forward limit. 
These bounds apply to any scalar EFT with a mass gap, which includes the massive Galileon on flat spacetime. The infinite number of bounds derived in  \cite{deRham:2017avq}  places constraints on the EFT coefficients, not only the leading Galileon operators, but also on all higher derivative operators that enter the massive Galileon LEEFT. \\

We begin by quickly reviewing the infinite number of scalar positivity bounds in section~\ref{sec:revspb} before focusing on the massive Galileon Wilsonian action in section~\ref{sec:Nonrenormalisation} which reviews a key similarity between the massive and massless Galileon, namely that they share a non-renormalization theorem which guarantees that only Galileon invariant local terms are generated in the 1PI effective action. We then apply the positivity bounds to the leading Galileon operators in section~\ref{sec:g3g4} and infer a bound on the coefficients that relates the cubic and the quartic Galileon operators. We then turn to the leading higher order derivative operators in section~\ref{sec:hd}, and derive their respective bounds.  We prove that no local, analytic  and Lorentz invariant Wilsonian UV completion could ever lead to the lowest order Galileon operators without also involving some higher derivative operators in the LEEFT, besides those being generically generated by quantum corrections,. To illustrate our results we provide in section~\ref{sec:UVmodel} a simple yet explicit example of a Lorentz invariant and renormalizable UV theory that leads to a special massive Galileon LEEFT in the sense that the LEEFT preserves the Galileon symmetry (up to the mass term) and enjoys the Galileon non-renormalization theorem. We conclude in section~\ref{sec:conc}. \\

We also supplement our results with four appendices: We generalize our results beyond the Galileon LEEFT to include the leading bounds for any massive and Lorentz invariant scalar LEEFT (without any assumption of symmetry) in Appendix~\ref{app:generalized}.
In appendix~\ref{app:positive} we prove that the positivity bounds are strictly positive definite (and can never be positive semi-definite in an interacting theory). In appendix~\ref{app:coeffs} we provide an alternative (but equivalent) formulation for the coefficients and functions that enter the positivity bounds. In appendix~\ref{app:diagonalize} we provide the explicit diagonalization used in the UV example provided in section~\ref{sec:UVmodel}.

\section{Review of the Scalar Positivity Bounds}
\label{sec:revspb}

In \cite{deRham:2017avq}, using the unitary and analytic properties of the scalar scattering amplitude, we have derived an infinite number of positivity bounds for the derivatives of pole subtracted amplitudes. In the following, we will review the main ingredients of the proof and results. The following discussion is not specific to Galileons, but we shall apply it to the massive Galileon LEEFT in sections~\ref{sec:g3g4}, \ref{sec:hd} and \ref{sec:UVmodel} and to a general massive scalar field LEEFT in Appendix~\ref{app:generalized}.

\subsection{Pole Subtracted Dispersion Relation}
In what follows we shall be interested in the $2\to 2$ scattering amplitude $A(s,t)$, for a single scalar species of mass $m$ which can be expanded into partial waves as
\be
A(s,t) = 16 \pi \sqrt{ \frac{s}{s-4m^2} } \sum^\infty_{\ell=0} (2\ell+1)P_\ell\left(\cos \theta \right) a_\ell(s)  ,
\ee
where $s,t,u$ are the usual Mandelstam variables and $\theta$ is the scattering angle  in the center of mass frame, $\cos \theta =1+\frac{2t}{s-4m^2}$. Making use of the partial wave unitarity bound ${\rm Im}\,a_\ell\geq |a_\ell|^2>0$ and the properties of the Legendre polynomial $P_\ell$, one can infer that in the physical region $s>4m^2$
\be
\frac{\pd^n}{\pd t^n} {\rm Im}A(s,t=0) \ge 0, ~~~{\rm for}~~~n\geq 0\,.
\ee
In appendix \ref{app:positive} we prove that taken together with the assumption of analyticity this is in fact a strict positivity
\be
\frac{\pd^n}{\pd t^n} {\rm Im}A(s,t=0) > 0, ~~~{\rm for}~~~n\geq 0\,.
\ee
By some weak assumptions on the analyticity on the Mandelstam complex plane, it has been shown \cite{Bros:1964iho,Martin:1965jj} that
$A(s,t)$ is analytic in the twice cut $s$ plane for fixed $t$ and in the disk $|t|<4m^2$ for fixed $s$ (excluding the obvious poles of $s$ and $t$)\footnote{The Mandelstam proposal would assume a much bigger analytic region.}.  This leads to
\be
\label{ImArel}
\frac{\pd^n}{\pd t^n} {\rm Im}A(s,t)>0, ~~~{\rm for}~~~n\geq 0 \text{ and for all } 0\leq t<4m^2\,, s\ge 4m^2\,.
\ee
On the other hand, making use of the Froissart-Martin bound \cite{Froissart:1961ux, Martin:1962rt,Jin:1964zza}, one can arrive at the twice subtracted dispersion relation
\be
\label{disrel}
A(s,t) = a(t) + \frac{\lambda}{m^2-s}
+\int^\infty_{4m^2} \frac{\ud \mu}{\pi} \frac{(s+\frac{t}{2}-2m^2)^2}{(\mu+\frac{t}{2}-2m^2)^2}\frac{{\rm Im}A(\mu,t)}{(\mu-s)}  + (s  \to u )\,,
\ee
where $a(t)$ is some unknown function and for scalar field theories $\lambda$ is independent of $s$ and $t$.
In what follows we shall be working in terms of the variable $v$ rather than the center of mass energy $s$,
\ba
 v=s+\frac{t}{2}-2m^2\,,
\ea
and denote by $B(v,t)$ the  pole subtracted dispersion relation
\ba
B(v,t)  = A(s,t)-\frac{\lambda}{m^2-s} -\frac{\lambda}{m^2-t} -\frac{\lambda}{m^2-u} \,.
\ea
In terms of $b(t)=a(t)-\lambda/(m^2-t)$, we have
\ba
 B(v,t)=b(t)+\int_{4m^2}^\infty \frac{2 \d \mu }{\pi}\frac{v^2}{(\mu+t/2-2m^2)}
\frac{{\rm Im}A(\mu,t)}{(\mu+t/2-2m^2)^2-v^2}\,.
\ea
The  derivatives of $B(v,t)$ are designated by $B^{(N,M)}$,
\be
B^{(N,M)}(t)  = \frac{1}{M!}
 \left. \pd^N_v \pd^M_t   B(v,t)\right|_{v=0}\,,
\ee
and are consequently being evaluated at $s = 2m^2-t/2$. Provided $ 0 \le t < 4m^2$, then $0 \le s \le 2m^2$, which lies in the Mandelstam triangle $0 \le s,t,u < 4m^2$ in which the pole subtracted amplitudes are known to be analytic, and hence all the derivatives are well defined.

\subsection{Positivity Bounds}
Combining the previous expressions together with the bound \eqref{ImArel} that follows from unitarity and analyticity,
we have shown in \cite{deRham:2017avq} that the following quantity must necessarily be positive definite if the theory is to have an analytic and Lorentz invariant UV completion,
\be
Y^{(2N,M)} (t) >0 ~~~ \text{for}~~~  N \ge 1 \, , \, M \ge 0  \, ,\quad 0  \le t < 4m^2  \,,
\ee
where $Y^{(2N,M)}(t)$ is defined by the following recurrence relation
\ba
 && \hspace{-15pt} Y^{(2N,0)} (t)  = B^{(2N,0)}(t)\,,\\
&& \hspace{-15pt}  Y^{(2N,M)} (t) = \sum_{r=0}^{M/2} c_r B^{(2(N+r),M-2r)}  + \frac{1}{{\cal M}^2} \sum_{ \text{even}\,k=0}^{(M-1)/2}  (2(N+k)+1) \beta_k   Y^{(2(N+k),M-2k-1)},\quad
\label{eqn:Y}
\ea
where ${\cal M}^2 = {\rm Min}( \mu + t /2-2m^2)=2m^2+t/2$,
and the coefficients $c_r$ and $\beta_k$ defined recursively by
\ba
\label{eq:coeffs}
c_k=-\sum_{r=0}^{k-1}\frac{2^{2(r-k)}c_r}{(2(k-r))!}\,, {\rm with }\quad c_0=1\,,
\quad
{\rm and }
\quad
\beta_k=(-1)^k \sum_{r=0}^k \frac{2^{2(r-k)-1}}{(2(k-r)+1)!}c_r \ge 0\,.\quad
\ea
Alternative but fully equivalent expressions for these coefficients are also given in appendix~\ref{app:coeffs}. In what follows we shall see how to apply those bounds to the massive Galileon LEEFT.

\subsection{Tree versus Loop Bounds}

\label{sec:Treeversusloops}

The bounds $Y^{(2N,M)} (t) >0$ derived in \cite{deRham:2017avq} are true for the full all-loop scattering amplitude. We can however, apply them directly to the tree level LEEFT. If we compute a scattering amplitude to tree level in the low energy effective theory, then there will be no imaginary parts in the region $4m^2 \le  \mu < \Lambda_{\rm th}^2$ where $ \Lambda_{\rm th}$ is the threshold to produce new heavy states, i.e. the mass of the lightest state that lies outside of the low energy effective field theory. This allows us to apply the bounds $Y_{\rm tree}^{(2N,M)} (t,\Lambda_{\rm th}) >0$  to the tree amplitude in which we take ${\cal M}^2 = {\rm Min}( \mu +  t /2-2m^2) = \Lambda_{\rm th}^2 -2m^2+t/2 \approx \Lambda_{\rm th}^2$.
In the application of the tree level bounds, we must however be careful in how to interpret the bounds on the higher derivative terms.
To see the problem let us first consider the forward scattering limit. If in a given theory the tree level scattering amplitude takes the form
\be
A_{\rm tree}(s,0) \sim c_1 \frac{s^2}{\Lambda^4} + c_2 \frac{s^4}{\Lambda^8} + \dots \, ,
\ee
while it is clear that the forward scattering limit bounds impose  $c_1>0$, we cannot further declare $c_2>0$ without further specifying how we separate trees and loops (the renormalization prescription), since on computing a one-loop diagram, we will obtain renormalization prescription sensitive local terms that will contribute at the same order $1/\Lambda^8$. The problem arises if there is a single scale in the problem, i.e. $\Lambda$, at which the theory is strongly coupled, then it is no longer possible to separate the tree and loop contributions that arise at the same order in a power expansion in $s/\Lambda^2$ since the loop expansion itself breaks down at $s \sim \Lambda^2$. \\

\paragraph{Weak coupling:} As noted in \cite{Adams:2006sv}, this problem is resolved if it is assumed that the UV theory is weakly coupled, in which case there exists an additional small parameter $g$ for which the tree scattering amplitude takes the form
\be
A_{\rm tree}(s,0) \sim g \left( \tilde c_1 \frac{s^2}{\Lambda^4} + \tilde c_2 \frac{s^4}{\Lambda^8} + \dots \) \, ,
\ee
then the one-loop contribution will be of order $g^2$ and so if $g \ll 1$ we can safely put the bound on all higher derivatives of $A_{\rm tree}$. So if we assume a weakly coupled UV completion, we can consistently impose $Y_{\rm tree}^{(2N,M)} (t,\Lambda_{\rm th}) >0$ for all $N$ and $M$. Interestingly, the explicit example we give for a UV completion of a massive Galileon in section \ref{sec:UVmodel} falls into this category.
\\

\paragraph{Massive Galileon:} In the case of the massive Galileon however we can do better. The tree amplitude for a massive Galileon takes the form
\be
A_{\rm tree}(s,\theta) \sim  \left( d_1(\theta) \frac{m^2 s^2}{\Lambda^6}+ d_2(\theta) \frac{s^3}{\Lambda^6} + d_3(\theta)\frac{s^4}{\Lambda^8}\dots  \)\,,
\ee
and the loop contributions\footnote{Here it is understood that all light loops are computed in dimensional regularization. Since all cutoff dependence can be absorbed into a redefintion of the tree level higher derivative operators, it is only necessary to track the running contributions. We take the point of view that the LEEFT Lagrangian is so defined.} begin at
\be
A_{\rm one-loop}(s,\theta) \sim \sum_{n=0}^3 \frac{\tilde d_n(\theta) m^{2 n} s^{6-2n}}{\Lambda^{12}} +\dots\,.
\ee
Given the assumption $m\ll \Lambda$, the loop corrections   to the tree level coefficients computed up to and including $1/\Lambda^{10}$ corrections are negligible relative to the existing tree level contribution. This holds even though the coefficient of $s^2$ is already $m^2/\Lambda^2$ suppressed. As a result we may take seriously all of our tree level bounds $Y_{\rm tree}^{(2N,M)} (t,\Lambda_{\rm th})>0 $ that will be computed in sections \ref{sec:g3g4} and \ref{sec:hd} applied up to and including order $1/\Lambda^{10}$, i.e. the finite number of them that receive only contributions from the tree level scattering amplitude expanded to this order.  This already allows us to put non-trivial bounds on the higher derivative coefficients without needing to assume that the UV completion is weakly coupled, and is precisely what we shall do in what follows. \\

\paragraph{Including Loops:} We can also go beyond this as follows. Given the same assumption of the hierarchy $m \ll \Lambda$, then to a given order in $1/\Lambda^2$ there are only a finite number of loops that contribute significantly. Specifically, if we compute the scattering amplitude to order $1/\Lambda^{2 K}$, then we need only compute loops to order $N_{\rm loop} = {\rm Floor}[K/3]$. We can then impose those loop level bounds $Y^{(2N,M)} (t) >0$ (with ${\cal M}^2 = {\rm Min}(\mu +  t /2-2m^2)=2m^2+t/2$), that include only contributions from the scattering amplitude coefficients up to order $1/\Lambda^{2K}$. These can be strengthened by using the knowledge of the light loops to that order, to compute their contribution to $B^{(N,M)}(t) $ in the region in which they can be computed perturbatively. This is achieved as follows: assuming that perturbation theory can be trusted up to a scale $\epsilon \Lambda \gg m $ where $\epsilon \ll 1$, then we may define (see \cite{deRham:2017avq} for the origin of this combination)
\be
B_{\epsilon \Lambda}^{(2N,M)}(t) = B^{(2N,M)}(t) - \sum_{k=0}^M \frac{2 (-1)^k}{\pi k! 2^k} \frac{(2N+k)!}{(M-k)!} \int_{4m^2}^{\epsilon^2 \Lambda^2} \d \mu \frac{\partial_t^{2N+k} {\rm Im} \, A(\mu,t)}{(\mu+t/2-2m^2)^{M-k+1}} \, . 
\ee
We can then compute $Y_{\epsilon \Lambda}^{(2N,M)}(t)$ out of $ B_{\epsilon \Lambda}^{(2N,M)}(t) $ via the recurrence relations defined in (\ref{eqn:Y}) where we now take ${\cal M}^2 =  \epsilon^2 \Lambda^2 + t /2-2m^2 \approx \epsilon^2 \Lambda^2$. Following the arguments of \cite{deRham:2017avq} we may then show that
\be
Y_{\epsilon \Lambda}^{(2N,M)}(t) >0 \, .
\ee
It is understood that if the amplitude is computed to order $1/\Lambda^{2K}$, then only those bounds $Y_{\epsilon \Lambda}^{(2N,M)}(t) >0 $ that include only contributions from derivatives that arise up to this order are taken seriously.  \\

The previous arguments ensure that for any LEEFT, we may apply the positivity bounds to any desired order in the EFT expansion, provided at least we include the light loops to the desired order. The strongest form of the bound will then be obtained by subtracting off the known contribution from the light loops to define $Y_{\epsilon \Lambda}^{(2N,M)}(t) $. In the case of an assumed weakly coupled UV completion, which our explicit example in section \ref{sec:UVmodel} falls into, we can impose all orders in $N,M$ of tree level bounds. In the special case of a massive Galileon, it is sufficient to work at tree level up to and including order $1/\Lambda^{10}$ which is what we do in the following.

%%%%%%%%%%%%%%%%
\section{Massive Galileon and Non-Renormalization Theorem}
\label{sec:Nonrenormalisation}
%%%%%%%%%%%%%%%%

\subsection{Massive Galileon Effective Field Theory}

In what follows we start by considering the Lagrangian for a massive extension to the Galileon \cite{Nicolis:2008in} which, in four flat spacetime dimensions,   takes the form
\ba
\L_{\text{mGal}}[\pi]&=& \mathcal{L}_{\rm Galileon}[\pi]  - \frac12 m^2 \pi^2  \\
 &=&  \sum_{n=2}^5 \frac{g_{n}}{n! \Lambda^{3(n-2)} } \;  \; \pi\,  \Pi^{\mu_1}_{[\mu_1} \cdots \Pi_{\mu_{n-1}]}^{\mu_{n-1}}    - \frac12 m^2 \pi^2     \label{eqn:Galileon} \\
&=& - \frac{1}{2} (\partial \pi )^2 - \frac12 m^2 \pi^2 +  \frac{g_3}{3! \Lambda^3} \pi\left[  [\Pi]^2 - [\Pi^2]     \right]\nn \\
&& +  \frac{g_4}{4! \Lambda^6} \pi\left[  [\Pi^3] - 3[\Pi][\Pi^2]  + 2 [\Pi^3]       \right]   +  \ldots\,, \nn
\ea
where we have used the notation $\Pi\mn=\p_\mu\p_\nu \pi$, square brackets represent the trace of a tensor, and antisymmetrization is defined without $1/n!$, e.g. $A_{[\mu \nu]}=A_{\mu \nu}-A_{\nu \mu}$. It is convenient to use the standard canonical normalization for the scalar field $g_2=1/2$. The dimensionless coefficient $g_3$ (or $g_4$) could be absorbed in the definition of $\Lambda$ but we keep it separate  for later convenience. ${\cal L}_{\rm Galileon}[\pi]$ contains the distinguished Galileon operators whose equations of motion are second order in derivatives.

Since the bounds derived in section~\ref{sec:revspb} only deal with the $2\to2$ scattering amplitude, and since we will mainly focus on tree-level for $\pi$, we do not need consider higher than quartic interactions.

\subsection{Wilsonian Action}

When viewed as an effective field theory, the Galileon must be supplemented by an infinite number of higher derivative operators which also respect the Galileon symmetry. As a result, the full Wilsonian action for this massive Galileon LEEFT $S_W[\pi] $ is then,
\be
\label{eq:SWil}
S_W[\pi] = \int \d^4 x \( \L_{\text{mGal}}[\pi] + \L_{\rm h.d.}(\p^2 \pi , \p^3 \pi,  \p^4 \pi, \dots) \) \, ,
\ee
where ${\cal L}_{\rm h.d.}$ is a scalar function constructed from all the possible tensor combinations of two or more derivatives acting on the field. The precise form of ${\cal L}_{\rm h.d.}$ depends strongly on the renormalization scheme employed to compute loops since these operators are expected to receive order unity renormalizations\footnote{We may for instance define $S_W$ in the Euclidean as the effective action in which trees and loops of the heavy fields for all momenta, and low energy field $\pi$ for momenta above the scale $\Lambda_{\rm th}$, are integrated out, so that the cutoff for the remaining light loop integrals is $\Lambda_{\rm th}$. However, a more practical definition is to define $S_W$ as the Wilsonian action for which the remaining $\pi$ loops can be computed in dimensional regularization. There is no loss in generality in this approach since the terms discarded in dimensional regularization are precisely the local operators already included in ${\cal L}_{\rm h.d.}$ }.\\

Stated differently, the massive Galileon may be defined as any effective field theory for which the action transforms under the transformation $\pi \rightarrow \pi + c+ v_{\mu} x^{\mu}$ as
\be
\delta_{c,v} S_W = - \int \d^4 x \,  m^2 \pi \, \( c+  v_{\mu} x^{\mu}\)  \,.
\ee
The massive Galileon has several important properties that put it on the same footing as its massless counterpart\footnote{For a recent discussion on non-renormalization theorems of this type see \cite{Goon:2016ihr}.}:
\begin{enumerate}

\item Quantum corrections preserve the Galileon symmetry (provided that the Galileon couples to all other fields through Galileon invariant interactions),

\item Quantum corrections do not renormalize the coefficients of the leading Galileon operators in $\mathcal{L}_{\rm mGal}$, \ie they neither renormalize the coefficients $g_n$, nor the mass scale $m$.

\end{enumerate}
To demonstrate this, it is simplest to consider the expression for the one-particle irreducible effective action $\Gamma[\pi]$, (see Ref.~\cite{Burrage:2011cr}),
\be
\exp \left[\frac{i}{\hbar} \Gamma[\pi]\right] = \int D \pi'  \exp\left[\frac{i}{\hbar}S_W[\pi'] - \frac{i}{\hbar} \frac{\delta \Gamma[\pi]}{\delta \pi}(\pi'-\pi)\right] \, .
\ee
The 1PI effective action can be split into its classical and quantum parts,
\be
 \Gamma[\pi] = S_W[\pi] + \hbar \, \Gamma_q[\pi] ,
\ee
and performing a similar split in the integration measure of the path integral $\pi' = \pi + \sqrt{\hbar} \chi$ we then have
\be
\exp{i \Gamma_q[\pi]} = \int D \chi  \exp\left[i F[\pi,\chi] - i \sqrt{\hbar}\frac{\delta \Gamma_q[\pi]}{\delta \pi}\chi\right] \, ,
\ee
where
\be
F[\pi,\chi]  = \sum_{n=2}^{\infty}  \hbar^{(n-2)/2} \frac{1}{n^!}\frac{\delta^n S_W[\pi]}{\delta \pi^n}  \chi^n \, .
\ee
As is well known the path integral may be evaluated to determine $\Gamma_q[\pi]$ as an expansion in powers of $\hbar$. The key observation is that although the addition of a mass term to $S_W$ breaks the Galilean symmetry of $S_W$, it nevertheless leaves invariant $\frac{\delta^n}{\delta \pi^n} S_W[\pi]$ for $n \ge 2$. This is manifest since
\be
\frac{\delta^2}{\delta \pi(x) \delta \pi(y) } S_W[\pi] = \frac{\delta^2}{\delta \pi(x) \delta \pi(y) } (S_{\rm Galileon}[\pi] + S_{\rm h.d.}[\pi] ) - m^2 \delta^4(x-y) \, ,
\ee
and for all higher functional derivatives the mass term does not enter, e.g.
\be
\frac{\delta^3}{\delta \pi(x) \delta \pi(y) \delta \pi(z)} S_W[\pi] = \frac{\delta^3}{\delta \pi(x) \delta \pi(y) \delta \pi(z)} (S_{\rm Galileon}[\pi] + S_{\rm h.d.}[\pi] )  \, .
\ee
We may thus conclude that $F[\pi]$ and hence $\Gamma_q[\pi]$ is manifestly Galileon invariant
\be
F[\pi ] = F[\pi+c + v_{\mu} x^{\mu}] \rightarrow \Gamma_q[\pi] = \Gamma_q[\pi+c + v_{\mu} x^{\mu}] \, .
\ee
This argument is easily generalized to include the Galileon interacting with other fields, provided that the interactions to the other fields are themselves manifestly invariant under the Galileon symmetry. \\

The second part of the non-renormalization theorem states that the Galileon operators $S_{\rm Galileon} $ and the mass term are not renormalized \cite{Luty:2003vm,Nicolis:2004qq}. This  follows simply from the observation that $\frac{\delta^2}{\delta \pi(x) \delta \pi(y) } S_W[\pi]$ and all higher derivatives depend on $\pi$ only in the combination of functions of $\Pi\mn$ and its derivatives.
For instance, for the leading massive Galileon operators
\be
\frac{\delta^2}{\delta \pi(x) \delta \pi(y) } S_{\rm mGal}[\pi] = \left[ \Box-m^2+ \sum^5_{n=3} \frac{g_n}{(n-2)!\Lambda^{3(n-2)}} \Pi^{\mu_1}_{[\mu_1}\cdots  \Pi^{\mu_{n-3}}_{\mu_{n-3}} \partial^{\mu_{n-2}} \partial_{\mu_{n-2}]}
 \right] \delta^4(x-y) \, ,
\ee
are explicitly only dependent on $\pi$ through powers of the invariant combination $\Pi\mn$. Similarly since $S_{\rm h.d.}$ itself only depends on functions of $\Pi\mn$ and its derivatives then the same clearly holds for all functional derivatives with respect to $\pi$.
This implies that the local counter terms generated in $\Gamma_q[\pi]$ will contain only local functions of $\Pi\mn$ and its derivatives. However, $S_{\rm mGal}[\pi]$ contains a dependence on $\pi$ through fewer than two derivatives. Thus no local counterterm generated in $\Gamma_q[\pi]$ can renormalize $S_{\rm mGal}[\pi]$, although they will renormalize $ S_{\rm h.d.}[\pi]$ which is precisely why the latter terms are included. Once again, the addition of a mass term does not affect this property since the mass only arises as a constant, $\pi$ independent term in $\frac{\delta^2}{\delta \pi(x) \delta \pi(y) } S_W[\pi] $. \\

\paragraph{Mass Parameter and Pole of the Propagator:}

A clear product of these results is that `mass' of the Galileon defined as the non-derivative coefficient of $\pi^2$ term in $\Gamma[\pi]$, \ie
\begin{equation*}
m^2 = - \left[ \int \d^4 x e^{ik.x}\delta^2 \Gamma/ \delta \pi(x) \delta \pi(0) \right]_{\pi=0,k=0}\,,
\end{equation*}
will also receive no quantum corrections. However the {\it physical mass}, defined as the pole of the propagator will receive finite quantum corrections from higher derivative terms. These occur because loops can generate terms in $\Gamma_q[\pi]$ of the Galileon invariant form $\pi \Box^{1+n} \pi/\Lambda^{2n}$ (with $n\ge 1$) which will shift the physical pole $m_{\rm phys}$ away from $m$,
\be
m_{\rm phys}^2 = m^2 \( 1 + \sum_{n=1}^{\infty} d_n \frac{m^{2n}}{\Lambda^{2n}}\)\,.
\ee
A small Galileon mass is technically natural as long as $m \ll \Lambda$, which is a manifestation of the t'Hooft naturalness argument. \\

In practice however, when computing the scattering matrix, it is always possible to ignore higher derivative contributions in $S_W$ to the quadratic action
\be
\Delta S_W = \int \d^4 x \sum_{n=1}^{\infty }  f_n \, \frac{\pi \Box^{1+n} \pi}{\Lambda^{2n}} \, ,
\ee
the reason being that all such terms can be removed by a field redefinition of the form
\be
\label{fieldredef1}
\pi \rightarrow \pi + \sum_{n=0}^{\infty }  f'_n \, \frac{\Box^{n}}{\Lambda^{2n}}  \pi \, ,
\ee
that has the virtue of preserving the Galileon symmetry. The field redefinition will have the affect of adjusting the coefficients in the Galileon invariant interactions and we can take the point of view that this has already been done from the outset.

%%%%%%%%%%%%%%%%
\section{Massive Galileon Leading Positivity Bounds}
\label{sec:g3g4}
%%%%%%%%%%%%%%%%

\subsection{Constraints on the Massive Galileon}

Now let us apply the positivity bound to the {\it tree} level $2\to2$ scattering amplitude of the massive Galileon.
Excluding the higher derivative operators, the $2\to 2$ scattering amplitude for the massive Galileon \eqref{eqn:Galileon} in the centre of mass frame is given by
\be
A(s,t) = A_s + A_t + A_u + A_4\,,
\ee
with
\ba
A_X
= \frac{g_3^2 X^2 \left( X-4 m^2\right)^2}{16 \Lambda^6 \left(m^2-X\right)}\,, \quad {\rm and }\quad
A_4 =
g_4 \frac{stu}{4 \Lambda^6}\,.
\ea
The pole subtracted amplitude is then
%\be
%B(s,t) = ( g_4 - \frac{3}{4} g_3^2 ) \frac{s t u}{4 \Lambda^6} + g_3^2 \frac{m^2 (s^2 + t^2 + u^2 ) }{ 16 \Lambda^6 } - \frac{31 g_3^2 m^6}{16 \Lambda^6}  ,
%\ee
\be
B (s,t) = a_{00}  + a_{10} x   +   a_{01} y ,
\label{eqn:anm}
\ee
which depends on $\{ s,t,u\}$ via the crossing symmetric combinations,
\ba
\label{eq:xy}
 x = - \left( \bar s \bar t + \bar t \bar u + \bar u \bar s \right)\,  , \qquad y = - \bar s \bar t \bar u
\ea
where the bar denotes $\bar s = s - 4m^2/3$, $\bar t = t-4m^2/3$, $\bar u = u-4m^2/3$. The coefficients are
\be
a_{00} =  \frac{m^6}{\Lambda^6} \left[ \frac{16 g_4}{27} - \frac{295g_3^2}{144}    \right] , \;\;\;\;
a_{10} =  \frac{m^2}{\Lambda^6} \left[  - \frac{g_4}{3} +\frac{3 g_3^2}{8}   \right], \;\;\;\;
a_{01} =   \frac{1}{{\Lambda^6}}\left[ - \frac{g_4}{4} + \frac{3 g_3^2}{16} \right] .
\ee
Significantly, for order unity $g_3$ and $g_4$, we see that $a_{10} \sim \tfrac{m^2}{\Lambda^2} a_{01} $. This is related to the fact that the massless Galileon has enhanced soft behaviour due to the Galileon symmetry \cite{Cheung:2014dqa,Cheung:2016drk}, and so in the limit $m \to 0$ we need $a_{10}$ to vanish.  \\

Note that these leading Galileon interactions \eqref{eqn:Galileon} have given a scattering amplitude truncated at sixth order in energy, and so from the results of \cite{Adams:2006sv, deRham:2017avq} we have two independent positivity bounds,
\begin{align}
\label{bd1}
Y^{(2,0)} &: ~~~  a_{10} + a_{01} {\bar t}  >  0
\\
\label{bd2}
Y^{(2,1)} &: ~~~  a_{01} + \frac{3 }{2 \Lambda_{\rm th}^2 } \left( a_{10} + a_{01} {\bar t} \right)    > 0\,,
\end{align}
where as  mentioned earlier, $\Lambda_{\rm th}$ is the threshold scale at which new physics must necessarily  enter to restore analyticity and unitarity (since we are dealing with the tree-level amplitude).  For the massive Galileon LEEFT to make sense at all, $\Lambda^2_{\rm th}$ should lie above the scale $4m^2$ and ideally around or above the scale $\Lambda^2$. \\

\begin{figure}
\centering
 \includegraphics[width=0.9\textwidth]{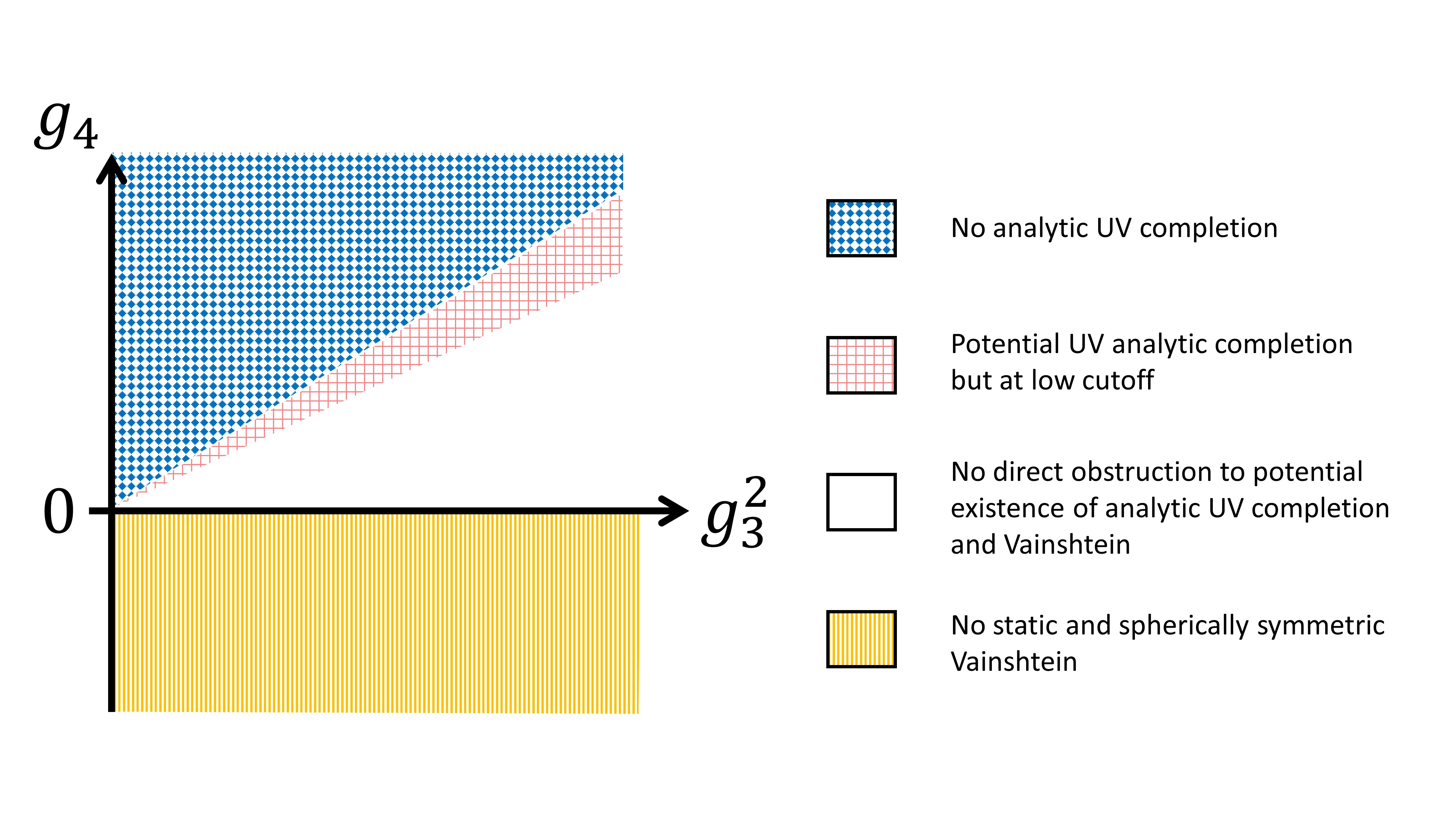}
%\raisebox{-0.5\height}{ \includegraphics[width=0.5\textwidth]{Lthm.pdf} }
\caption{Summary of constraints imposed on coefficients of cubic ($g_3$) and quartic ($g_4$) Galileon operators to respect known  bounds imposed by (1). the existence of a hypothetical
local, analytic UV completion, (2). a cutoff which is above the Galileon mass, and (3). the existence of a static and spherically symmetric Vainshtein mechanism. The boundary between no analytic Wilsonian UV completion and a potential UV completion with an unacceptably low cutoff is at $g_4=7/8 g_3^2$. For $g_4<3/4 g_3^2$ there is to date no known obstruction for the potential existence of an analytic UV completion.   \label{fig:g3g4}}
\end{figure}

\noindent We can distinguish between 3 different scenarios, depending on the ratio $g_4/g_3^2$,
\begin{enumerate}
\item If $g_4/g_3^2 \le 3/4$, then both bounds (\ref{bd1}, \ref{bd2}) are satisfied for any value of $0 \le t <4m^2$, and without any restriction neither on $\Lambda_{\rm th}$, nor on the mass (this implies that the Galileon mass can be taken to be arbitrarily small, without violation of these bounds).
\item For $3/4 < g_4/g_3^2\le 7/8$, analyticity imposes the following upper bound on $\Lambda_{\rm th}$,
    \ba
    \label{eqn:g3g4}
    \Lambda_{\rm th}^2< 4m^2 \frac{\frac{7}{8}-\frac{g_4}{g_3^2}}{\frac{g_4}{g_3^2}-\frac 34 }\,.
    \ea
    Interestingly the strongest form of this bound arises not in the forward scattering limit $t\to 0$, but rather in the opposite limit $t\to 4m^2$ and the above upper bound corresponds to $t=4m^2$. This illustrates the power of extending the constraints beyond the forward scattering limit.
\item For $g_4/g_3^2> 7/8$, the Galileon model can enjoy no local, analytic and Lorentz invariant UV completion.
\end{enumerate}
Even though  the ratio $g_4/g_3^2$ can in theory be larger than $3/4$,
in practise, since the LEEFT only makes sense if $\Lambda_{\rm th}^2 \gg m^2$, the ratio $g_4/g_3^2$ can never get much larger than $3/4$.
The allowed region of parameter space is shown in figure~\ref{fig:g3g4}.
We emphasise that there is no condition imposed on $g_5$, which does not contribute to the $2\to2$ tree-level amplitude.

%%%%%%%%%%%%%%%%%%%%%%%%%%
%%%%%%%%%%%%%%%%%%%%%%%%%

\subsection{Strong Coupling Scale}

It is interesting to relate this to the scale for which perturbative unitarity breaks down for $2\to 2$.  This occurs when any of the partial waves violate the optical theorem, $| a_{\ell} (s) | < 1$. In this case, the largest multipole moment is at $\ell = 0$,
\bal
32 \pi \sqrt{ \frac{s}{s-4m^2} } a_0 (s) &=  \int_{-1}^1 \d \cos \theta \; P_0 (\cos \theta) A (s,t)
\\
&=  (3 g_3^2 - 4g_4) \frac{s^3}{24\Lambda^6} - \left(g_3^2 -2 g_4 \right)\frac{2  s^2m^2}{3 \Lambda ^6}  +  \mathcal{O} \(\frac{s m^4}{\Lambda^6} \) \,,
\eal
focusing again on tree-level contributions since our aim here is to compare with the bounds obtained previously from tree-level $2\to 2$ scattering.
Then generically the strong coupling scale implied by this process is
\be
\Lambda_{\text{strong coupling}} = \frac{\Lambda}{|g_4-3 g_3^2/4|^{1/6}}\,,
\ee
unless we artificially tune $ (g_4-3 g_3^2/4)$ to be small (\ie of order $m^2/\Lambda^2$ or smaller). This combination of parameters has a natural explanation. In the absence of a mass term, we can use the Galileon duality transformation \cite{deRham:2013hsa, deRham:2014lqa} to remove the cubic interaction and put it in the quartic interaction (and quintic one, which is irrelevant to this discussion). The combination $g_4'=g_4-3 g_3^2/4$ is precisely the new coefficient of the quartic Galileon operator after this transformation has been performed. Thus tuning $g_4-3 g_3^2/4=g_4'$ to be small is in effect artificially making the strong coupling scale (related to tree-level $2\to2$ scattering) large by switching off interactions. Given this it is more natural to define $\Lambda$ (which has so far remained a free parameter) as the strong coupling scale in the $m \rightarrow 0$ limit, which at the level of the $2\to 2$ scattering amplitude amounts to setting $|g_4-3g_3^2/4|=1$. With this convention we see that the bound on $\Lambda_{\rm th}$ in the region $4/3 < g_4/g_3^2\le 7/8$ is
\be
\Lambda_{\rm th}^2 < \frac{1}{2} m^2 g_3^2
\ee
and with the usual reasonable assumption that $g_3$ is of order unity we would find at best $\Lambda_{\rm th}  \sim m$, which renders the LEEFT inconsistent. This is an example of where, as pointed out in \cite{deRham:2017avq}, it is possible that analyticity acts as a stronger constraint on the cutoff of an effective field theory than perturbative unitarity alone. \\

Taking into account these points, the bounds from analyticity at this order effectively imply that
\be
g_4/g_3^2 \le 3/4 \, , \quad  \text{ \ie } \quad g_4'=g_4-3 g_3^2/4 <0 \, .
\ee

\subsection{Analyticity vs Vainshtein Mechanism}

\label{sec:Vainshtein}

At this stage it is interesting to compare how these bounds overlap with the requirement that the Galileon exhibits a Vainshtein mechanism \cite{Nicolis:2008in}. Although not central to their use as effective field theories, for the majority of phenomenological applications it is necessary that Galileons are in the Vainshtein screened region which suppresses their contributions to fifth forces, evading otherwise strong solar system constraints on gravity.  \\

For instance, for a spherically symmetric source, such as a star, the Vainshtein mechanism requires that we can find a real regular solution to the static spherically symmetric Galileon equations.  When working with the leading Galileon operators ${\cal L}_{\rm Galilon}$, since the quintic Galileon vanishes in 3d, \ie vanishes in any static configuration, then the quintic Galileon (or the coefficient $g_5$) does not enter this argument. From \cite{Nicolis:2008in}, see eqns.(57-60), we see that the Vainshtein only works in that static and spherically symmetric configuration if
\be
g_3>-\sqrt{g_4} \, , \quad  g_4 \ge 0 \, .
\ee
In the special case where $g_3=0$, the positivity bound Eqn.~\eqref{eqn:g3g4} requires that
\be
g_4=g_4' <  0 \, ,
\ee
which then excludes any possibility that a low energy effective field theory could have a unitary and analytic UV completion, with an active phenomenological Vainshtein mechanism.
More generally the combined requirements imply
\be
g_4>0 \,,  \quad \quad g_3 > \sqrt{\frac{4 g_4}{3}}  >0 \, .
\ee

%%%%%%%%%%%%%%%%
\section{Positivity bounds on Higher Derivative Terms}
\label{sec:hd}
%%%%%%%%%%%%%%%%

Having shown that the massive Galileon is consistent with the leading order positivity constraints, we may look to higher order bounds $Y^{(2N,M)}>0$. Based on the tree level computation done so far involving the operators in $\L_{\rm mGal}$, we have that $B^{(2N,M)} = 0$ for all $ 2N + M \geq 4$, which automatically violates the higher order positivity bounds. Whilst loops from the light fields will generate a non-zero contribution to these $B^{(2N,M)}$, the tree level bounds are putting constraints on the unknown heavy physics which UV completes the Galileon. This heavy physics which has been integrated out gives rise to higher derivative operators in the EFT, previously denoted as ${\cal L}_{\rm h.d.}$ in \eqref{eq:SWil}, and will contribute to the tree level  amplitudes and to $B^{(2N,M)}$. Consequently analyticity and unitarity of the unknown UV physics will impose constraints on the coefficients of these higher derivative operators.\\

For instance, to account for the leading order contributions to the $2\to2$ scattering amplitudes from this heavy physics (\ien, ${\cal O}(1/\Lambda^8)$), we must include the following cubic and quartic higher derivative interactions
\ba
{\cal L}_{\rm h.d.} &=& \frac{1}{\Lambda^5} \left( c_1 [\Pi^3]+ c_2   [\Pi^2]  [\Pi]+ c_3   [\Pi]^3 \right) \nn \\
&+& \frac{1}{\Lambda^8 } \left( d_1   [\Pi^4] + d_2   [\Pi^3]  [\Pi]+ d_3   [ \Pi^2]  [ \Pi]^2    + d_4  [\Pi^2]^2+d_5 [\Pi]^4 ) \right) \, .
\ea
Each of these terms will give rise to contributions to the $2\to2$ scattering that scale as $1/\Lambda^8$. This is clear for the quartic interactions, and for the cubic it arises for diagrams for which one vertex is a $1/\Lambda^3$ interaction (from the cubic Galileon)  and the second is a $1/\Lambda^5$ interaction. Hence on dimensional grounds alone, these tree level interactions will give a contribution to scattering amplitude of the form $s^4/\Lambda^8$ which will show up as a finite contribution to the higher order $Y^{(2N,M)}, \; 2N + M = 4$ bounds. Although at the level of the Lagrangian we seem to have 8 undetermined coefficients at this order in derivatives, they are actually related by various field redefinition redundancies and total derivatives (for instance the $[\Pi]^3$ is actually equivalent to $3[\Pi][\Pi^2]-2[\Pi^3]$, and a similar relation holds for $[\Pi]^4$).\\

Explicitly, the pole subtracted scattering amplitude is
\begin{equation}
B (s,t) = a_{00}  + a_{10} x   +   a_{01} y + a_{20} x^2 \,,
\end{equation}
where $x$ and $y$ are expressed in terms of the Mandelstam variables in \eqref{eq:xy},
$x = - \left( \bar s \bar t + \bar t \bar u + \bar u \bar s \right)$ and $y = - \bar s \bar t \bar u$. The coefficients in the expression of $B(s,t)$ are
\begin{align}
a_{10} &= \frac{m^2}{\Lambda^6} \left[ - \frac{g_4}{3} + \frac{3g_3^2}{8}   \right] + \frac{m^4}{\Lambda^8} \left[ 2 d_2 + 4 d_3 + \frac83 d_4 - \frac{g_3}{12} \left( 57 c_1 + 14 c_2 - 72 c_3  \right)    \right]  \, , \\
a_{01} &= \frac{1}{\Lambda^6} \left[ - \frac{g_4}{4} + \frac{3 g_3^2}{16} \right]  + \frac{m^2}{ \Lambda^8} \left[ -2 d_1 - 3 d_2 + 4 d_4 - \frac{g_3}{8} (3c_1 + 2 c_2 ) \right] \, ,   \\
a_{20} &=   \frac{1}{\Lambda^8} \left[\frac{d_1}{2} + d_4 + \frac{g_3}{4} ( 3 c_1 + 2 c_2) \right] \,    .
\end{align}
These are bounded by,
\begin{align}
Y^{(2,0)} &: ~~~ a_{10} + a_{01} {\bar t} + \frac{3 }{2 } a_{20} \bar t^2 >  0 \, ,   \\
Y^{(2,1)} &: ~~~ a_{01} + 3 a_{20} \bar t   + \frac{3}{2 \Lambda_{\rm th}^2 } \left[ a_{10} + a_{01} {\bar t} + \frac{3 }{2 } a_{20} \bar t^2 \right]  > 0 \, , \\
Y^{(4,0)} &:~~~ a_{20}  >  0, .
\end{align}
These are the only independent bounds at this order. The higher order bounds with $2N+M>4$ cannot be computed without a knowledge of the amplitude beyond $O(1/\Lambda^8)$. The first 2 bounds are the same as the bounds of Eq (\ref{bd1}) and (\ref{bd2}), but now include small corrections of ${\cal O}(1/\Lambda^8)$ from higher order derivative terms.  The bound of $Y^{(4,0)} $ yields
\be
 \label{eqn:LthL}
\frac{d_1}{2} + d_4 + \frac{g_3}{4} \left( 3 c_1 + 2 c_2 \right)    > 0\,.
\ee
This bound is of course easily satisfied as we have (superficially) 4 new parameters that enter unsuppressed. In practice not all of these parameters are independent because of the ability to do field redefinitions, however the combination $\frac{d_1}{2} + d_4 + \frac{g_3}{4} \left( 3 c_1 + 2 c_2 \right)$ is automatically invariant under field redefinitions. Crucially it is not possible to set $d_1=d_4=c_1=c_2=0$. \\

Thus the existence of a local UV completion requires that the LEEFT has non-zero higher derivative operators. From an EFT point of view this is not too surprising since these operators will inevitably be generated from loops of the heavy fields. The new input is that already at tree level it is necessary to include these operators, i.e. it is not possible to tune the theory so that all these higher derivative terms vanish at some scale.  They necessarily arise from integrating out the heavy fields that UV complete the theory. Once again, if light loops are computed in dimensional regularization, then they will make only $(m/\Lambda)^4$ suppressed contributions to the coefficients at this order.\\

This trend will continue if we look at higher order contributions to the amplitude, as more indices will come in and thus more possible operators at each order. For example, up to ${\cal O}(1/\Lambda^{10})$ contribution, we need to include operators that schematically are of the form
\be
{\cal L}_{\rm h.d.} \supset \frac{1}{\Lambda^7} \partial^2 \Pi^3
+ \frac{1}{\Lambda^{10} } \partial^2 \Pi^4\,,
\ee
At each new order we will obtain new bounds, but the increase in the number of new coefficients will adequately compensate this. As we have already discussed in section \ref{sec:Treeversusloops}, once we reach the order $1/\Lambda^{12}$ then in the absence of a weak coupling parameter the one-loop of the light field contribute at the same order. The higher loops remain suppressed as long as $m \ll \Lambda$. It is then necessary to either apply the exact version of the bounds, or follow the method discussed in section \ref{sec:Treeversusloops} and impose the bounds $Y_{\epsilon \Lambda}^{(2N,M)}(t) >0$.

\section{UV Completion: A Simple Example}
\label{sec:UVmodel}

\subsection{Manifestly Galileon Invariant Formulation}

In this section, we consider a simple UV completion of a massive Galileon, obtained via the introduction of a single heavy field $H$ of mass $M_H$.  After integrating out the heavy field, we obtain manifestly Galileon invariant interactions for the light Galileon field $\pi$.
Remembering that a massive Galileon can be defined as a theory for which under the Galileon transformation $\pi \rightarrow \pi + v_{\mu}x^{\mu}$ the Lagrangian transforms as ${\delta {\cal L}} = -m^2 \pi v_{\mu} x^{\mu}$ up to total derivatives, then it is straightforward to see that the following renormalizable theory respects this symmetry:
\be
\label{actUVgal}
S_{\rm UV}[\pi,H]= \int \ud^4 x \left ( -\frac12 (\pd \pi)^2  - \frac12 (\pd H)^2 - \ai  H \Box \pi   - \frac12 m^2 \pi^2 - \frac12 M_H^2 H^2 - \frac{\lambda}{4!} H^4 \right)  \,   .
\ee
Here we require $|\ai|<1$ to avoid a ghost instability and $ M_H\gg  m$ to set an appropriate EFT hierarchy that allows us to integrate out the heavy field.
The Wilsonian effective action for the massive Galileon is defined via the path integral
\be
e^{i S_W[\pi]}  = \int \D H \, e^{i S_{\rm UV}[\pi,H] } \, ,
\ee
and will take the form of an expansion in loops of the heavy field
\be
S_W = \sum_{n=0}^{\infty} \, S_W^{(n)} \, ,
\ee
where $n$ counts the number of heavy loops (\ie of loops of the heavy field $H$). \\

Explicitly integrating out the heavy field $H$ to determine $S^{(0)}_W$ corresponds to solving the classical equation of motion for $H$ to give $H_{\rm tree}$, and then substituting back in the Lagrangian. This leads to
\bal
S^{(0)}_W &= \int \ud^4 x \bigg[ -\frac12 (\pd \pi)^2  - \frac12 m^2 \pi^2 +\frac{\ai^2}{2M_H^2}\pi\Box^2 \pi + \frac{\ai^2}{2M_H^4} \pi \Box^3\pi + \frac{\ai^2}{2M_H^6}  \pi \Box^4\pi
\nn\\
&~~~~~~~~~~~~~~~~~~~~~+ \frac{\ai^2}{2M_H^8}  \pi \Box^5\pi -\frac{\li\ai^4}{4!}\frac{(\Box\pi)^4}{M_H^8} + {\cal O}\left(\frac{1}{M_H^{10}}\right) \bigg]  \,  .
\eal
As mentioned above, the higher derivative quadratic terms can be removed by a field redefinition at the price of redefining the coefficients of the interactions. To this order however the resulting interactions are relatively uninteresting since for example the operator $(\Box \pi)^4$, although Galileon invariant, can be field redefined into $m^8 \pi^4+\dots$, and so at tree level in the heavy fields there are no truly higher derivative interactions. \\

The situation is different if we include loops from the heavy fields. For instance, at one-loop the action picks up a contribution
\be
S^{(1)}_W = \int \d^4 x  \, {\cal L}^{(1)}_W = - \frac{1}{2}{\rm Tr} \ln [\Box -M_H^2 - \lambda H_{\rm tree}^2(x)] \, .
\ee
Expanding this we will, for example, obtain terms of the form
\be
   {\cal L}^{(1)}_W  \supset \frac{\lambda^2}{M_H^{2n}} H^{2}_{\rm tree} \Box^n H^2_{\rm tree} \, ,
 \ee
and since $H_{\rm tree}$ depends on $\pi$, $H_{\rm tree} \sim - \alpha \Box \pi/M_H^2 + \dots$ then this corresponds to interactions
\be
 {\cal L}^{(1)}_W \supset \frac{\alpha^4 \lambda^2}{M_H^{8+2n}} (\Box \pi)^2 \Box^n[(\Box \pi)^2] \, .
\ee
Once again, these interactions are manifestly Galileon invariant, as required by the non-renormalization theorem, and can be field redefined into
\be
  {\cal L}^{(1)}_W  \supset \frac{\alpha^4 \lambda^2 m^8 }{M_H^{8+2n}} \pi^2 \Box^n \pi^2+\dots \, ,
\ee
corresponding to genuinely non-trivial higher derivative interactions. These show up in the scattering amplitude as $s$ dependent contributions
\be
A(s,\theta)  \supset  \kappa_n(\theta) \frac{\alpha^4 \lambda^2 m^8 }{M_H^{8+2n}} s^n \, ,
\ee
which we will see explicitly in the exact form of the scattering amplitude given below.

\subsection{Diagonalized Formulation}

In practice, to calculate the scattering amplitude in the UV theory, it is easier to work with an action in which both the kinetic term and mass terms are diagonalized:
\be
S_{\rm UV}= \int \ud^4 x \left ( -\frac12 (\pd \tilde\pi)^2  - \frac12 (\pd \tilde H)^2 - \frac12 \tilde m^2 \tilde\pi^2 - \frac12 \tilde M^2 \tilde H^2 - \frac{\tilde\lambda}{4!} (\tilde H+\beta \tilde \pi)^4 \right) \, .
\ee
The explicit form of the diagonalization transformations is given in appendix \ref{app:diagonalize}. The salient point is that for $M_H \gg m$, $\tilde m \sim m$, $\tilde M^2 \sim M_H^2/(1-\alpha)$, $\tilde \pi = \pi - \alpha H$, $\tilde H \sim \sqrt{1-\alpha^2} H$, $\beta \sim - \alpha \sqrt{1-\alpha^2} m^2/M_H^2$. In this form the Galileon symmetry is realized in the sense
\ba
&& \tilde \pi \rightarrow \tilde \pi + v_{\mu} x^{\mu}  \, ,\\
&& \tilde H \rightarrow \tilde H - \beta v_{\mu} x^{\mu} \, ,  \\
&& {\cal L} \rightarrow {\cal L} - \( \tilde m^2 \pi - \beta  \tilde M^2 \tilde H  \) v_{\mu} x^{\mu}   \, .
\ea
Even though the heavy field shifts in this representation, since the shift is linear, the non-renormalization theorem remains unaffected. \\

%%%%%%%%%%%%%%%%%%%%%%%%%%
\begin{figure}
\centering
 \includegraphics[width=\textwidth]{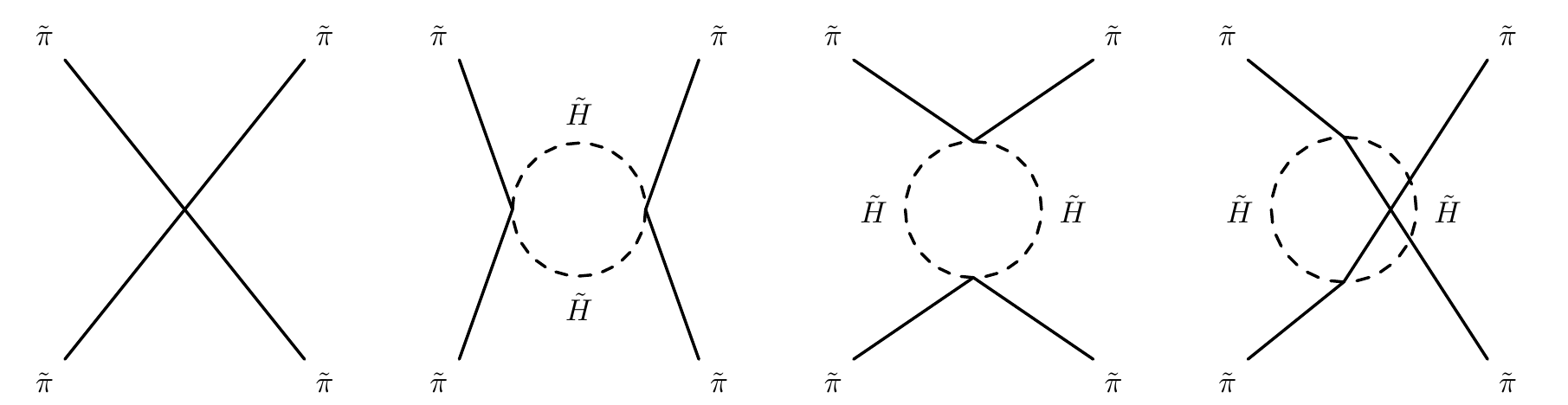}
\caption{$\tilde \pi \tilde \pi \to \tilde \pi \tilde \pi$ diagrams up to 1-loop in the heavy field. We emphasize that from the LEEFT picture these are all tree-level diagrams of the light field. }
\label{fig:feyn}
\end{figure}
%%%%%%%%%%%%%%%%%%%%%%%%%

We are interested in calculating the $2\to2$ scattering amplitude between the light fields $\tilde\pi\tilde\pi \to \tilde\pi\tilde\pi$. According to Eq (\ref{tpitopi}), the $\pi$ and $\tilde\pi$ fields are slightly different, but as a result of the equivalence theorem and the LSZ formalism the S-matrix for the two sets of asymptotic states evaluated on-shell are the same.
Up to one loop in the heavy field, we only have the diagrams given in Fig.~\ref{fig:feyn} for the $2\to2$ scattering of $\tilde\pi$. The amplitude is given by
\be
A = A_{4} + A_s+A_t+A_u \, ,
\ee
where
\ba
&& A_{4}  = - \tilde\li \beta^4 \, ,
\\
&& A_X  = -\frac{\tilde\li^2\bi^4}{32\pi^2}\int^1_0\ud x \ln \frac{\tilde M^2-X x (1-x)}{\mu^2} \, ,
\ea
computed using dimensional regularization ($\tilde \li \rightarrow \mu^{4-d}\tilde\li$) in the $\overline{\rm MS}$ subtraction scheme. \\

The closed form of the integral $A_X$ depends on the value of $X$. To make use of our positivity bounds, we can focus on the range $0\leq X<4\tilde m^2\ll 4\tilde M^2$, within which we have
\be
A_X = -\frac{\tilde\li^2\bi^4}{32\pi^2} \left[ \ln\frac{\tilde M^2}{\mu ^2}-2+2 \sqrt{\frac{4 \tilde M^2-X}{X}} {\rm arccsc}\left( \frac{2\tilde M}{\sqrt{X}}\right) \right] \, .
\ee
Due to the absence of cubic interactions, the amplitude $A$ does not have poles at the mass $\tilde m^2$, so we have
\be
B= A_{\rm tree} + A_s+A_t+A_u \, .
\ee
Note that as argued in \cite{deRham:2017avq}, only the light field loops contribute to the imaginary part of the amplitude ${\rm Im} A(\mu,t)$ when $\mu$ lies in the range $4\tilde m^2$ to $\tilde M^2$. So for our positivity bounds, we may choose ${\cal M}^2\sim \tilde M^2$.  In this explicit example it happens that there are also no tree level contribution from the heavy field $H$ and so the real threshold for new physics is $4\tilde M^2$. So, explicitly, we can choose  ${\cal M}^2 = (8\tilde M^2-4\tilde m^2 +t)/2$ in calculating $Y^{2N,M}(t)$. Up to the leading $t$ dependence, the first few positivity bounds are
\bal
Y^{2,0}(t) & =\frac{\tilde\li^2\bi^4}{32\pi^2} \frac{1}{\tilde M^4}\left[ \frac{1}{15}+ \frac{ \left(4 \tilde m^2-t\right)}{70 \tilde M^2} + {\cal O}\(\frac{\tilde m^4}{\tilde M^4}\) \right]>0 \, ,
\\
Y^{2,1}(t) & =\frac{\tilde\li^2\bi^4}{32\pi^2} \frac{1}{\tilde M^6} \left[ \frac{3}{280 } + \frac{5 \left(4 {\tilde m}^2-t\right)}{1344 {\tilde M}^2}+ {\cal O}\(\frac{\tilde m^4}{\tilde M^4}\)  \right]>0\, ,
\\
Y^{4,0}(t) &= \frac{\tilde\li^2\bi^4}{32\pi^2} \frac{1}{\tilde M^8}\left[ \frac{2}{105 } + \frac{2 \left(4 \tilde m^2-t\right)}{231 \tilde M^{2}}+ {\cal O}\(\frac{\tilde m^4}{\tilde M^4}\)  \right]>0\, ,
\\
Y^{4,1}(t) & =\frac{\tilde\li^2\bi^4}{32\pi^2} \frac{1}{\tilde M^{10}}\left[ \frac{1}{308 } + \frac{61 \left(4 {\tilde m}^2-t\right)}{32032 {\tilde M}^{2}}+ {\cal O}\(\frac{\tilde m^4}{\tilde M^4}\) \right]>0\, ,
\\
Y^{4,2}(t) & = \frac{\tilde\li^2\bi^4}{32\pi^2} \frac{1}{\tilde M^{12}}\left[\frac{5}{2464 } + \frac{185 \left(4 {\tilde m}^2-t\right)}{128128 {\tilde M}^{2}}+ {\cal O}\(\frac{\tilde m^4}{\tilde M^4}\) \right]>0\, ,
\\
Y^{6,0}(t) & = \frac{\tilde\li^2\bi^4}{32\pi^2} \frac{1}{\tilde M^{12}} \left[ \frac{20}{1001 } +\frac{2 \left(4 {\tilde m}^2-t\right)}{143 {\tilde M}^{2}}+ {\cal O}\(\frac{\tilde m^4}{\tilde M^4}\)\right]>0\, ,
%\\
%Y^{6,1}(t) & = \frac{\tilde\li^2\bi^4}{32\pi^2} \left[ \frac{1}{286 {\tilde M}^{14}}+\frac{113 \left(4 {\tilde m}^2-t\right)}{38896 {\tilde M}^{16}}+ {\cal O}(\tilde M^{-18}) \right]>0\, ,
%\\
%Y^{6,2}(t) & = \frac{\tilde\li^2\bi^4}{32\pi^2} \left[ \frac{7}{2288 {\tilde M}^{16}} + \frac{35 \left(4 {\tilde m}^2-t\right)}{11968 {\tilde M}^{18}} + {\cal O}(\tilde M^{-20}) \right] >0\, ,
%\\
%Y^{6,3}(t) & =  \frac{\tilde\li^2\bi^4}{32\pi^2} \left[\frac{37331}{5912192 {\tilde M}^{18}}+ \frac{321177 \left(4 {\tilde m}^2-t\right)}{47297536 {\tilde M}^{20}}+ {\cal O}(\tilde M^{-22}) \right]>0\, ,
\\
\vdots &{~~~~~~~~~~~} \vdots \nn
\eal
which are manifestly positive definite for $0\leq t<4m^2$, as required. Higher orders are just proportional to powers of $(4m^2-t)$ which are negligible given the assumed hierarchy $\tilde M \gg \tilde m$. In this example, we see this to be a weakly coupled UV completion of the massive Galileon, along the lines discussed in section \ref{sec:Treeversusloops} with the small parameter
\be
g \sim \bi^4 \ll 1\, .
\ee
For example at one-loop in the light field, we will have a term with two $A_{\rm tree}  = - \tilde\li \beta^4$ vertices coming in at order $\tilde \li^2 \beta^8$ which is suppressed by $\beta^4$ relative to the heavy loop.

\subsection{Massless limit}

Having given an explicit UV completion of a massive Galileon, it is interesting to explore how this is consistent with known properties of the massless limit. In this concrete example, the cutoff of the low energy effective theory is the mass of the heavy particle $\tilde M  \approx M_H$. In a standard massless Galileon theory in which all the coefficients are of order $\Lambda$, the leading term in the scattering amplitude is of the form
\be
A \sim \frac{(s^3+t^3+u^3)}{\Lambda^6} \, ,
\ee
whereas by contrast in this UV completion the analogous coefficient at this order is
\be
A \sim \lambda^2 \frac{m^8}{M_H^8} \frac{\alpha^4}{M_H^6} (s^3+t^3+u^3) \, .
\ee
Identifying the two we see that $\Lambda \sim M_H (M_H/m)^{4/3} \li^{-1/3}\alpha^{-2/3}$ and so in the limit $m \rightarrow 0$, $\Lambda \rightarrow \infty$. In other words, the massless limit of our massive Galileon UV completion, is not a massless Galileon but simply a free theory. This is transparent from the action \ref{actUVgal} where in the limit $m\rightarrow 0$ we may redefine $\pi = \hat \pi - \alpha H$ to give an interacting heavy field and a decoupled free scalar $\hat \pi$:
\be
\lim_{m \rightarrow 0} S_{\rm UV}[\pi,H]= \int \ud^4 x \left ( -\frac12 (\pd \hat \pi)^2  - \frac12 (1-\alpha^2 )(\pd H)^2    - \frac12 M_H^2 H^2 - \frac{\lambda}{4!} H^4 \right)  \,   .
\ee
Stating this differently, in the limit $m \rightarrow 0$, keeping $\Lambda$ fixed, the cutoff of the Galileon EFT $M_H \ll \Lambda$ tends to zero. In this way we are not in conflict with the statement of \cite{Adams:2006sv} that the massless Galileon does not have an analytic UV completion.

\subsection{Weak Coupling UV Completions}

It is possible to argue quite generally that if the massive Galileon has an analytic UV completion, and if it is not strongly coupled, then it becomes a free theory in the limit $m \rightarrow 0$  \cite{Bellazzini:2016xrt}, as in the above example. Let us assume that the threshold for new physics is some heavy mass $M_H$, and introduce a weak coupling parameter $g$ so that the tree level massive Galileon scattering amplitude takes the form
\be
A_{\rm tree}(s,\theta) \sim g \left( d_1(\theta) \frac{m^2 s^2}{M_H^6} + d_2(\theta) \frac{s^3}{M_H^6}  +d_3(\theta) \frac{s^4}{M_H^8} + \dots \right) \, .
\ee
Here $d_2(\theta)$ are the usual Galileon interactions that already arise in the $m=0$ limit, $d_1(\theta)$ are the corrections that arise when a mass is included and $d_3(\theta) + \dots $ come from the higher derivative operators that we have seen must necessarily be included.
Loop corrections will kick in at 
\be
A_{\rm one-loop}(s,\theta)  \sim g^2 \frac{s^6}{M_H^{12}} \tilde d_0(\theta) + \dots \, , 
\ee
and the regime of validity of perturbation theory is expected to be $\sqrt{s} \ll M_H/g^{1/6}$. If the theory is weakly coupled $g \ll 1$ then we can trust the perturbative expansion all the way up to $M_H$. \\

Focussing on the leading forward limit bound $B^{(2,0)}(0) >0$ we have
\be
\partial_s^2 B(s=2m^2,t=0) = \int_{4m^2}^{\infty} \frac{4 \d \mu}{\pi} \frac{{\rm Im} A(\mu,0)}{(\mu-2m^2)^3} >0  \, .
\ee
Separating out the light and heavy parts of the integrals and using the positivity of the integrand for all $\mu \ge 4m^2$ as a result of the optical theorem we also have
\be
\partial_s^2 B (s=2m^2,t=0) > \int_{4m^2}^{M_H^2} \frac{4 \d \mu}{\pi} \frac{{\rm Im} A(\mu,0)}{(\mu-2m^2)^3}  \, .
\ee
Using the scattering amplitude to one loop order, then evaluating the left and right hand sides for $m \ll M_H$ this approximates to
\be
\frac{g m^2}{M_H^6} \gtrsim g^2 \frac{M_H^{8}}{M_H^{12}} \, \quad \Rightarrow \quad 
g \lesssim \frac{m^2}{M_H^2} \, .
\ee
We thus conclude that since $m \ll M_H$, any such UV completion will be weakly coupled $g \ll 1$ and that in the massless limit $m \rightarrow 0 $, the theory becomes free $g \rightarrow 0$ \cite{Bellazzini:2016xrt}. \\

Defining the usual Galileon scale $\Lambda$ by comparing the coefficients of $s^3$ in the tree amplitude then we have
\be
\Lambda^6 = M_H^6/g \gtrsim \frac{M_H^8}{m^2} \,  , \quad \Rightarrow \quad M_H \lesssim (m^2 \Lambda^6)^{1/8} \, . 
\ee
The tree level LEEFT Lagrangian for such a weakly coupled UV completion will take the form
\ba
\L_W[\pi] &=&  - \frac{1}{2} (\partial  \pi )^2 - \frac12 m^2  \pi^2 +  \frac{ g_3}{3! \Lambda^3} \pi \left[  [ \Pi]^2 - [ \Pi^2]     \right]  +  \frac{ g_4}{4! \Lambda^6} \pi\left[  [ \Pi^3] - 3[\Pi][ \Pi^2]  + 2 [ \Pi^3]    + \dots   \right]   \nn \\
&& +  \frac{\Lambda^6}{M_H^2} \tilde \L_{\rm h.d.}\(\frac{\p^2  \pi}{\Lambda^3} , \frac{\p^3  \pi}{M_H \Lambda^3},  \frac{\p^4 \pi}{M_H^2 \Lambda^3}, \dots \)    \,, 
\ea
where $ g_n$ are order unity coefficients and $\tilde \L_{\rm h.d.}$ is a dimensionless scalar function of all contractions of its arguments with order unity coefficients. Now we see in order to be in a region where the leading Galileon operators dominate the classical solution, the gradients of the classical field configurations need to satisfy $\partial \ll M_H $. Nevertheless, from this argument the weakly coupled UV completion is not in conflict with the possibility of a Vainshtein mechanism, provided the bounds derived in section \ref{sec:Vainshtein} are satisfied and the gradients are under control.
%%%%%%%%%%%%%%%%
\section{Discussion}
\label{sec:conc}
%%%%%%%%%%%%%%%%

Since its first appearance within the context of (soft) massive gravity theories, the consistency of Galileon LEEFTs has remained a matter of much debate over the past decade \cite{Adams:2006sv}.
 While a mass would technically break the Galileon symmetry it does so in a way that preserves the non-renormalization theorem and all the essential features  of the Galileon. Moreover in  most of its known realizations, \ie within the context of massive gravity theories, the Galileon appears as the helicity-0 mode of the graviton in a particular decoupling limit and {\it is massive} away from that decoupling limit. It is therefore natural to include the mass as part of the Galileon LEEFT. \\

In parallel, assuming the existence of {\it any} local, Lorentz-invariant and analytic Wilsonian UV completion imposes an infinite number of positivity bounds on the $2\to 2$ scattering amplitude and its derivatives of any scalar LEEFT with a mass gap, \cite{Adams:2006sv,deRham:2017avq} and are hence directly applicable to the massive Galileon LEEFT. Using {\it all the tree-level} positivity bounds known so far (both those previously derived in the literature as well as the new ones very recently derived in \cite{deRham:2017avq}), we have shown the existence of an entire region of parameter space which shows no obstruction (at tree-level) to the potential existence of a standard Wilsonian UV completion.
A direct consequence of the positivity bounds derived in \cite{deRham:2017avq} is that higher derivative operators (that also respect the Galileon symmetry) are necessarily required to be present if the LEEFT is to have a standard Wilsonian UV completion. \\

Further requiring the existence of an active Vainshtein mechanism for static and spherically symmetric configuration does reduce this region but not entirely. However we emphasize that the analysis performed in this work has nothing to say about the validity of the Vainshtein regime where the field is strongly coupled (see \cite{deRham:2014wfa} for a discussion on this point). \\

The absence of direct obstructions to the potential existence  of a standard Wilsonian UV completion from $2\to 2$ tree-level considerations alone, are by no means to be taken as an indication that such a UV completion will definitively exist.  In the context of the leading Galileon operators (that do not get renormalized), finding such a UV completion would certainly be a success in itself, which is of course well-beyond the scope of this work. However for a particular massive Galileon LEEFT we were able to provide an explicit example of UV completion, which illustrates the fact that a Wilsonian UV completion is indeed possible and even explicitly constructible in some of these massive Galileon LEEFT.  \\

In the context of the specific UV complete example we have found, we can manifestly see that in the massless limit, the Galileon LEEFT either becomes a trivial free theory or its cutoff vanishes. This realization is fully consistent with the results found in \cite{Adams:2006sv} arguing for the absence of standard Wilsonian UV completion  for the massless Galileon. For that case, alternatives to the usual Wilsonian picture seem to remain as the only possibility  \cite{Dvali:2010jz,Dvali:2010ns,Dvali:2011nj,Dvali:2011th,Vikman:2012bx,Kovner:2012yi,Keltner:2015xda}.

\vskip 10pt
\acknowledgments
We would like to thank Brando Bellazzini for useful comments. CdR thanks the Royal Society for support at ICL through a Wolfson Research Merit Award. SM is funded by the Imperial College President's Fellowship. AJT thanks the Royal Society for support at ICL through a Wolfson Research Merit Award.

\appendix

%%%%%%%%%%%%%%%%
\section{General Massive Scalar LEEFT}
\label{app:generalized}
%%%%%%%%%%%%%%%%

The operators which provide leading order contributions (up to $\mathcal{O} \left( \Lambda^{-6} \right)$) to the four point function in the most general massive scalar LEEFT (deprived of any particular symmetry) are,
\begin{align}
\mathcal{L}[\phi] &= - \frac{1}{2} ( \partial \phi )^2 - \frac{1}{2} m^2 \phi^2 + m c_{30}  \phi^3  + \frac{c_{32}}{\Lambda}\phi (\partial \phi )^2  + \frac{g_{3}}{3! \Lambda^3} \phi \, \partial_{[\mu_1} \partial^{\nu_1} \phi \; \partial_{\mu_2]} \partial^{\nu_2} \phi    \label{eqn:Lfull} \\
&+ c_{40} \, \phi^4 + \frac{ c_{42}}{ \Lambda^2 }  \phi^2 (\partial \phi )^2 + \frac{c_{44}}{\Lambda^4} (\partial \phi )^4 + \frac{g_{4}}{4! \Lambda^6} \phi \partial_{[\mu_1} \partial^{\nu_1} \phi \; \partial_{\mu_2} \partial^{\nu_2} \phi \; \partial_{\mu_3]} \partial^{\nu_3} \phi   \,, \nonumber
\end{align}
up to total derivatives, where $X_{[\mu \nu]} = X_{\mu \nu} - X_{\nu \mu}$. In principle, one might also have added the following additional interactions: $(\Box \phi)^3$, $(\Box \phi)(\p_\mu \p_\nu \phi)^2$,  $\phi (\Box \phi)^2$, $\phi  (\partial \phi )^2 \Box \phi$, $\phi (\Box \phi)^3$ and $\phi (\Box \phi) (\p_\mu \p_\nu \phi)^2$, however as the leading order equations of motion relate $\Box \phi$ to $m^2 \phi$, we are guaranteed to have a field redefinition which replaces these operators by the ones already included in \eqref{eqn:Lfull}.
In this formulation  $(c_{nm}, g_n)$ represent 7 undetermined EFT coefficients, and $\Lambda$ is an arbitrary scale introduced to make them dimensionless. \\

We note in passing that this is the same theory that one would obtain by expanding the generalized Galileon \cite{Deffayet:2011gz},
\begin{align}
\mathcal{L} [\phi] = \sum_{n=0}^d A_n \left( \phi, (\partial \phi)^2 \right) \partial^{\mu_1} \partial_{[\mu_1} \phi ... \partial^{\mu_n} \partial_{\mu_n ]} \phi
\end{align}
to this order, where $A_n$ are independent analytic functions of $\phi$ and $(\partial \phi)^2$. However this is largely a coincidence, there is no reason for \eqref{eqn:Lfull} to agree with the generalized Galileon at higher orders (without some degree of fine tuning).
\\

The most general 2-to-2 scattering amplitude for a single scalar field, up to $\mathcal{O} \left( \Lambda^{-6} \right)$, is then given from \eqref{eqn:Lfull} as,
\begin{equation}
A (s,t)  =  A_s +  A_t +  A_u + A_4
\label{eqn:A}
\end{equation}
\begin{align}
A_X &=  \frac{1}{m^2 -X} \left[ 6 m c_{30}  - c_{32} \frac{X+2m^2}{\Lambda} - \frac{g_{3}}{4} \frac{X (X-4m^2) }{\Lambda^3}   \right]^2  ,  \\
A_4 &=   24 c_{40} - 8 \frac{m^2}{\Lambda^2} c_{42} + 2 c_{44} \frac{s^2 +t^2 + u^2 - 4m^4}{\Lambda^4} + \frac{g_4}{4} \frac{stu}{\Lambda^6}  .
\end{align}
This gives rise to the pole subtracted amplitude
\be
B (s,t) = a_{00}  + a_{10} x   +   a_{01} y ,
\ee
with
\begin{align}
a_{00} &=  24 c_{40} + \frac{36 m}{\Lambda} c_{30} c_{32} - \frac{m^2}{\Lambda^2} \left( 19 c_{32}^2 + 8 c_{42}  \right)   + \mathcal{O} \left( \frac{m^3}{\Lambda^3} \right) ,  \\
a_{10} &=  \frac{1}{\Lambda^4} \left[  4 c_{44} - c_{32} g_3  \right] +   \frac{m^2}{\Lambda^6} \left[  - \frac{g_4}{3} +\frac{3 g_3^2}{8}   \right],  \\
a_{01} &=  \frac{1}{\Lambda^6} \left[    - \frac{g_4}{4} + \frac{3 g_3^2}{16}   \right]  .
\end{align}
Note that $a_{10}$ is no longer $m^2/\Lambda^2$ suppressed, which means that for this case the $t$ dependence of the positivity bounds can be viewed as a small effect, as for $0 \leq t < 4m^2$ this dependence is suppressed by $m^2/\Lambda^2$. Truncating the amplitude to ${\cal O}(\Lambda^{-6})$, the $Y^{(2,0)}(\bar t=0)$ and $Y^{(2,1)}(\bar t=0)$ bounds are respectively
\begin{align}
 & 4 c_{44} - c_{32} g_3  > 0  \\
& \frac{ \Lambda^2 }{\Lambda_{\rm th}^2 } >  \frac{g_4 - \tfrac{3}{4} g_3^2  }{ 6 ( 4 c_{44} -  c_{32} g_3 ) }
\end{align}
Including a nonzero $c_{44}$ or $c_{32}$ makes it much easier to satisfy the positivity bounds for a wide range of $( g_3, g_4 )$. This is not surprising, the source of tension between Galileon theories and positivity has always been that Galileon symmetry seemed to prevent operators from contributing to $B^{(2,0)}$, and so discarding the Galileon symmetry naturally eases this tension. \\

Although these bounds are far fewer in number than the coefficients  in the effective Lagrangian, this is a reflection of the fact that many of these operators are redundant to this order since they may be removed by further field redefinition. For example,
\begin{equation}
\begin{split}
\phi \to \phi &+ \frac{d_{20}}{\Lambda} \phi^2 + \frac{d_{30}}{\Lambda^2} \phi^3 + \frac{d_{22}}{\Lambda^3} (\partial \phi)^2 + \frac{d_{32}}{\Lambda^4} \phi (\partial \phi)^2  +  \frac{ 2 d_{22}^2 }{\Lambda^6} \, \partial_\mu \phi \partial^\mu \partial^\nu \phi \partial_\nu \phi   + ...
\label{eqn:redef}
\end{split}
\end{equation}
where we will work to order $\phi^4$ only. Performing this field redefinition on \eqref{eqn:Lfull} we find that, at this order, it transforms into itself with modified coefficients,
\begin{align}
c_{30}' &= c_{30}  -  d_{20} \frac{m}{\Lambda}  \\
c_{32}' &= c_{32} - 2 d_{20} - d_{22} \frac{m^2}{\Lambda^2}   \\
g_{3}' &= g_{3}  - 4 d_{22} \\
c_{40}' &= c_{40}  + \frac{3m}{\Lambda} c_{30} \, d_{20}   - \frac{m^2}{2 \Lambda^2} \left( 2 d_{30} + d_{20}^2 \right)   \\
c_{42}' &= c_{42}  - 3 d_{30} - 2 d_{20}^2   + 5 c_{32} \, d_{20} + \frac{3m}{\Lambda} c_{30} \, d_{22}  - \frac{m^2}{\Lambda^2} \left(  d_{32} +  d_{20} d_{22} \right)     \\
c_{44}^{'} &= c_{44}  + 2 d_{20} d_{22} - c_{32} d_{22} - \frac{1}{2} g_{3} d_{20} + \frac{m^2}{2 \Lambda^2} d_{22}^2  \\
g_{4}' &= g_{4}   + 12 d_{22}^2   - 6 g_{3} \, d_{22}
\end{align}
again up to total derivatives and $\Box \phi$ operates which can be removed by a further redefinition.
This transformation preserves $S$ matrix elements, and indeed we find that the amplitude \eqref{eqn:A} is invariant. A special case of transformations of this form is the Galileon duality \cite{deRham:2013hsa,} which has the additional property that it forms a group, two duality transformations taken consecutively are equivalent to a single duality transformation. The ability to perform field redefinitions represents a degeneracy in our EFT parameters, seemingly different $(c_{nm}, g_n)$ are describing identical theories. This degeneracy can be removed by fixing a choice of the four coefficients $d_{nm}$ in \eqref{eqn:redef}. For example, one can use $d_{20}$ and $d_{22}$ to set $c_{32} = c_{44} = 0$, and then $d_{30}$ or $d_{32}$ to set $c_{42} = 0$. This leaves us with the massive Galileon \eqref{eqn:Galileon}, plus a $\pi^3$ and $\pi^4$ vertex, which do not contribute to the large $s$ behaviour of the amplitude, and hence do not affect the unitarity of a UV completion. Indeed, with this choice of $d_{mn}$, we find that the positivity bounds reduce to those found in the main text \eqref{eqn:g3g4}.  \\

Therefore for any scalar field theory on flat space, regardless of whether or not it has Galileon symmetry, the leading order positivity bounds on the four point function can be written as the bounds \eqref{eqn:g3g4} after an appropriate field redefinition.

\section{Proof that $\partial_t^n {\rm Im}A(s,0) > 0$ in the physical region.}
\label{app:positive}

In this appendix we prove that while the optical theorem implies a semi-definite bound on the imaginary part of the coefficients to the partial wave expansion of the $2\to2$ scattering amplitude, we necessarily have a definitive positive bound on the imaginary part of the amplitude and its derivatives, rather than a semi-definite bound. \\

First, it is straightforward to show from the partial wave expansion that $\partial_t^n {\rm Im}A(s,0) \ge 0$, for $s\ge 4m^2$. This just follows from the fact that
\be
\partial_t^n {\rm Im}A(s,0)  = 16 \pi \sqrt{\frac{s}{s-4m^2}}  \frac{2^n}{(s-4m^2)^n}\sum_{\ell=n}^{\infty} (2\ell+1) P_\ell^{n}(1) {\rm Im}(a_\ell(s)) \, ,
\ee
together with $P_\ell^{n}(1)= \partial_t^n P_\ell(1+t) |_{t=0}>0$, and ${\rm Im}(a_\ell(s)) \ge 0$ for $s$ in the physical region $s\ge 4m^2$. Furthermore it is clear for $n=0$ that since at least one of the ${\rm Im}(a_\ell(s))$ must be nonzero we are not dealing with a trivial free theory that ${\rm Im}A(s,0)>0$.\\

A priori, from considerations of unitary alone, it is not possible however to exclude the possibility that there could exist some $n_*$ for which $\partial_t^{n_*} {\rm Im}A(s,0)  =0$. This can only be achieved by imposing ${\rm Im}(a_\ell(s))=0$ for $\ell\ge n_*$, which in turn implies $\partial_t^n {\rm Im}A(s,0)=0$ for $n \ge n_*$. This means that the scattering amplitude only contains a finite number of partial waves, something which seems physically improbable but is not excluded by unitarity alone. \\

Fortunately this possibility can be ruled out based on the assumption of analyticity. Since the partial waves satisfy ${\rm Im}(a_l(s)) \ge |a_l|^2$, then we infer that we would require $a_l(s) =0$ for $l \ge n_*$ which in turn implies $\partial_t^n A(s,0) =0$ for all $n\ge n_*$.
Let us assume that this were the case, and then consider the twice subtracted dispersion relation
\ba
A(s,t) &=& a(t) + \frac{\lambda}{m^2-s}+ \frac{\lambda}{-3m^2 +t +s}  \nn \\
&+& s^2 \int_{4m^2}^{\infty} \frac{\d \mu}{\pi} \frac{{\rm Im} A(\mu,t)}{\mu^2 (\mu-s)} + (4m^2-t-s)^2 \int_{4m^2}^{\infty} \frac{\d \mu}{\pi} \frac{{\rm Im} A(\mu,t)}{\mu^2 (\mu-4m^2+t+s)} \, .
\ea
Differentiating twice, we get
\ba
\partial_s^2 A(s,t) &= &  \frac{2\lambda}{(m^2-s)^3}+ \frac{2\lambda}{(-3m^2 +t +s)^3}   \nn \\
&+& 2 \int_{4m^2}^{\infty} \frac{\d \mu}{\pi} \frac{{\rm Im} A(\mu,t)}{(\mu-s)^3} +2 \int_{4m^2}^{\infty} \frac{\d \mu}{\pi} \frac{{\rm Im} A(\mu,t)}{ (\mu-4m^2+t+s)^3} \, ,
\ea
then assuming that $\partial_t^{n_*} {\rm Im} A(s,0)  =0$ we have
\ba
\hspace{-5pt} && \partial_t^{n_*}\partial_s^2 A(s,t) =  \frac{(2+n_*)!}{2!} \frac{2(-1)^{n_*}\lambda}{(-3m^2 +t +s)^{3+n_*}}  \nn \\
&+& 2 \sum_{m=0}^{n_*-1} \int_{4m^2}^{\infty} \frac{\d \mu}{\pi} \frac{(-1)^{n_*-m} \partial_t^m {\rm Im} A(\mu,t)}{ (\mu-4m^2+t+s)^{3+n_*-m}} \frac{n_*! (2+n_*-m)!}{2! m!(n_*-m)!}  + \dots\, ,
\ea
where $\dots$ denotes terms which vanish at $t=0$ and whose $t$ derivatives vanish at $t=0$.
If we act on this with the operator $\partial_t - \partial_s $, this acts on terms in the denominators to give zero, and so we infer that
\ba
\hspace{-20pt} (-1)^{n_*} \( \partial_t - \partial_s\)^{n_*-1} \partial_t^{n_*}\partial_s^2 A(s,t)  &=& 2  \int_{4m^2}^{\infty} \frac{\d \mu}{\pi} \frac{\partial_t^{n_*-1} {\rm Im} A(\mu,t)}{ (\mu-4m^2+t+s)^{3+n_*}} \frac{(2+n_*)!}{2!(n_*)!} + \dots   \, ,
\ea
where by assumption $n_*$ is the lowest value of $n$ for which $\partial_t^n {\rm Im}(A(s,0))=0$ so that there is some range of $\mu \ge 4m^2$ for which $\partial_t^{n_*-1} {\rm Im} A(\mu,0) >0$.
Finally, evaluating this expression at $t=0$ we infer a contradiction:
\be
0 > 0 \, .
\ee
This then invalidates our initial assumption, implying that there is at least some range of $s$ for which
\be
\partial_t^n {\rm Im}A(s,0) >0 \, , \quad  s \ge 4m^2 \, ,  \quad  \forall \quad n\ge 0 \, .
\ee
Thus locality (analyticity) requires that an interacting theory has interactions for arbitrarily large partial waves.

\section{Equivalent Expressions for the Positivity Bound}
\label{app:coeffs}

In \cite{deRham:2017avq}, we have proven that the combinations  $Y^{(2N,M)}(t)$ defined as follows
\ba
Y^{(2N,0)} (t)  &=& B^{(2N,0)}(t)\,,\\
Y^{(2N,M)} (t)  &=& \sum_{r=0}^{M/2} c_r B^{(2(N+r),M-2r)}  + \frac{1}{{\cal M}^2} \sum_{ \text{even}\,k=0}^{(M-1)/2}  (2(N+k)+1) \beta_k   Y^{(2(N+k),M-2k-1)},\quad
\label{eqn:Y2}
\ea
satisfy a positivity bound $Y^{(2N,M)}(t)>0$ that follows from analytic and unitary considerations.
The coefficients $c_r$ and $\beta_k$ can be given recursively by
\ba
\label{eq:coeffs2}
c_k=-\sum_{r=0}^{k-1}\frac{2^{2(r-k)}c_r}{(2(k-r))!}\,, {\rm with }\quad c_0=1\,,
\quad
{\rm and }
\quad
\beta_k=(-1)^k \sum_{r=0}^k \frac{2^{2(r-k)-1}}{(2(k-r)+1)!}c_r \ge 0\,.\quad
\ea
 These  coefficients can also be expressed in terms of the Euler numbers $E_{2k}$  and Bernoulli   numbers $B_{k}$,
 \be
 c_k = \frac{E_{2k}}{(2k)!2^{2k}},~~~~~\beta_k= (-1)^k  \frac{( 2^{2k+3} -2)B_{2k+2}}{(2k+2)!}   \,.
 \ee
 Alternatively, we may also notice that these coefficients are simply the coefficients of the Taylor expansion of ${\rm sech}$ and ${\rm tan}$:
\be
{\rm sech}(x/2) = \sum_{k=0}^{\infty}  c_k x^{2k} ~~~ \text{and} ~~~
{\rm tan}(x/2) = \sum_{k=0}^{\infty} \beta_k x^{2k+1}    .
\ee
This allows the recursive definition of $Y^{(2N,M)}$ to be solved solely in terms of $\tilde B(v,t)$,
\begin{equation}
 Y^{(2N,M)} (t) = \frac{1}{M!} \partial_{v}^{2N} \partial_{t'}^M \left[  \hat{\mathcal{D}}_{2N} \left( t' -t , (t' -t ) \partial_v \right)  \,\tilde B (v, t') \right] \Big|_{v=0, t'=t}\,,
\end{equation}
 As a function, the function $\mathcal{D}$ can be defined as follows,
\begin{equation}
\mathcal{D}_{2N} ( t , x ) =  \text{sech} \left( \frac{x}{2} \right)  \frac{1}{1 - \frac{t}{\mathcal{M}^2} \left[ (2N+1) F(x) + x F' (x) + x F (x) \partial_x   \right]  }  \,1
\end{equation}
with $F(x) =  [\text{tanh} (x/2) + \text{tan} ( x/2 ) ]/(2x) $, but as an operator, $\hat{\mathcal{D}}_{2N}$ is to be understood as the  Taylor series expansion in $t$ of the previous function,
\begin{align}
\hat{\mathcal{D}}_{2N} ( t, t \partial_v ) &= 1 +  t\frac{ (2N+1) }{2 \mathcal{M}^2}  + t^2 \left( \frac{(2N+1)^2}{4 \mathcal{M}^4} - \frac{1}{8} \partial_{v}^2  \right) + t^3 \left(  \frac{ (2N+1)^3}{8 \mathcal{M}^6} - \frac{2N+1}{16 \mathcal{M}^2} \partial_v^2    \right)
\nn\\
&+ t^4 \left(  \frac{(2N+1)^4}{16 \mathcal{M}^8} - \frac{(2N+1)^2}{32 \mathcal{M}^4} \partial_v^2   + \frac{5}{384} \partial_{v}^4  \right) + \ldots\,.
\end{align}
From these relations,  one can read off the $Y^{(2N,M)}$ bounds. In terms of the original Mandelstam variable, we have
\begin{equation}
 Y^{(2N,M)} (t) = \frac{1}{M!}\,  \partial_{s}^{2N} \! \left( \partial_{t'} - \frac{1}{2} \partial_s  \right)^{\!M}  \left[  \mathcal{D}_{2N} \left( t' -t , (t' -t ) \partial_s \right)  \; B (s, t') \right] \Big|_{s=(4m^2-t)/2, t'=t}  > 0 .
\end{equation}

\section{Diagonalization}
\label{app:diagonalize}

The explicit form of field redefinitions that diagonalizes the kinetic and mass terms for our simple example UV completion $S_{\rm UV}$ introduced in section~\ref{sec:UVmodel} are
\ba
&& \tilde \pi = \frac{M_2}{2 \sqrt{2} \alpha  M_H^2 M_1} \left[(M_1^2+M^2-m^2) \pi -2 \alpha M_H^2 H\right] \, ,
\\
&& \quad = (\pi -\alpha  H)+\frac{\alpha  \left(\alpha ^2-1\right) H m^2}{M_H^2}+ H {\cal O}\(\frac{m^4}{M_H^4}\) \, ,
\\
&& \tilde H =   \frac{M_3}{2 \sqrt{2} \alpha  M_H^2 M_1} \left[(M_1^2-M_H^2+m^2) \pi +2 \alpha M_H^2 H\right] \, ,
\\
&& \quad = \sqrt{1-\alpha ^2} H-\frac{\alpha  \sqrt{1-\alpha ^2} m^2 (\alpha  H-\pi )}{M_H^2} + H {\cal O}\(\frac{m^4}{M_H^4}\)  \, ,
\ea
where
\ba
 && M_1^4 = 4 \alpha ^2 m^2 M_H^2+\left(M_H^2-m^2\right)^2 \,  ,
\\
&& M_2^2 =M_1^2+m^2+\left(2 \alpha ^2-1\right) M_H^2  \,,
\\
&& M_3^2 = M_1^2-m^2-\left(2 \alpha ^2-1\right) M_H^2 \, , \\
&& \bi = \frac{M_3 \left(M_H^2-m^2-M_1^2\right)}{M_2 \left(M_H^2-m^2+M_1^2\right)} =-\frac{\alpha  \sqrt{1-\alpha ^2} m^2}{M_H^2}  + {\cal O}(m^4) .
\ea
The natural mass and coupling constants in the redefined Lagrangian are
\ba
&& \tilde m^2= \frac{2 m^2 M_H^2}{m^2+M_H^2+M_1^2} = m^2\(1 + {\cal O}\(\frac{m^2}{M_H^2}\)  \)\, ,
\\
&&\tilde M^2= \frac{2 m^2 M_H^2}{m^2+M_H^2-M_1^2} = \frac{M_H^2}{1-\alpha ^2}+ \frac{\alpha ^2 m^2}{1-\alpha ^2} + {\cal O}\(\frac{m^4}{M_H^2}\)  \, ,
\\
&& \tilde \lambda = \lambda \left(\frac{M_1^2+M_H^2-m^2}{\sqrt{2} M_1 M_3}\right)^4 \, .
\label{tpitopi}
\ea

%\vspace{3cm}

%%%%%%%%%%%%%%%%
\bibliographystyle{JHEP}
\bibliography{references}

\providecommand{\href}[2]{#2}\begingroup\raggedright\begin{thebibliography}{10}

\bibitem{Nicolis:2004qq}
A.~Nicolis and R.~Rattazzi, {\it {Classical and quantum consistency of the DGP
  model}},  {\em JHEP} {\bf 06} (2004) 059,
  [\href{http://xxx.lanl.gov/abs/hep-th/0404159}{{\tt hep-th/0404159}}].

\bibitem{Dvali:2000hr}
G.~Dvali, G.~Gabadadze, and M.~Porrati, {\it {4-D gravity on a brane in 5-D
  Minkowski space}},  {\em Phys.Lett.} {\bf B485} (2000) 208--214,
  [\href{http://xxx.lanl.gov/abs/hep-th/0005016}{{\tt hep-th/0005016}}].

\bibitem{Luty:2003vm}
M.~A. Luty, M.~Porrati, and R.~Rattazzi, {\it {Strong interactions and
  stability in the DGP model}},  {\em JHEP} {\bf 09} (2003) 029,
  [\href{http://xxx.lanl.gov/abs/hep-th/0303116}{{\tt hep-th/0303116}}].

\bibitem{Adams:2006sv}
A.~Adams, N.~Arkani-Hamed, S.~Dubovsky, A.~Nicolis, and R.~Rattazzi, {\it
  {Causality, analyticity and an IR obstruction to UV completion}},  {\em JHEP}
  {\bf 0610} (2006) 014, [\href{http://xxx.lanl.gov/abs/hep-th/0602178}{{\tt
  hep-th/0602178}}].

\bibitem{deRham:2017avq}
C.~de~Rham, S.~Melville, A.~J. Tolley, and S.-Y. Zhou, {\it {Positivity Bounds
  for Scalar Theories}},  \href{http://xxx.lanl.gov/abs/1702.0613}{{\tt
  arXiv:1702.0613}}.

\bibitem{Eden:2012}
R.~J. Eden, P.~V. Landshoff, D.~I. Olive, and J.~C. Polkinghorne, {\em {The
  Analytic S-Matrix}}.
\newblock Cambridge University Press, 2002.

\bibitem{Nicolis:2008in}
A.~Nicolis, R.~Rattazzi, and E.~Trincherini, {\it {The Galileon as a local
  modification of gravity}},  {\em Phys. Rev.} {\bf D79} (2009) 064036,
  [\href{http://xxx.lanl.gov/abs/0811.2197}{{\tt arXiv:0811.2197}}].

\bibitem{Cheung:2014dqa}
C.~Cheung, K.~Kampf, J.~Novotny, and J.~Trnka, {\it {Effective Field Theories
  from Soft Limits of Scattering Amplitudes}},  {\em Phys. Rev. Lett.} {\bf
  114} (2015), no.~22 221602, [\href{http://xxx.lanl.gov/abs/1412.4095}{{\tt
  arXiv:1412.4095}}].

\bibitem{Cheung:2016drk}
C.~Cheung, K.~Kampf, J.~Novotny, C.-H. Shen, and J.~Trnka, {\it {A Periodic
  Table of Effective Field Theories}},
  \href{http://xxx.lanl.gov/abs/1611.0313}{{\tt arXiv:1611.0313}}.

\bibitem{Azimov:2011nk}
{\relax Ya}.~Azimov, {\it {How Robust is the Froissart Bound?}},  {\em Phys.
  Rev.} {\bf D84} (2011) 056012, [\href{http://xxx.lanl.gov/abs/1104.5314}{{\tt
  arXiv:1104.5314}}].

\bibitem{Diez:2013masters}
V.~Diez, {\it Improvements to froissart bound from ads/cft correspondence},
  Master's thesis, KTH, School of Engineering Sciences, 2013.
\newblock DiVA: diva2:648093.

\bibitem{deRham:2009rm}
C.~de~Rham, {\it {Massive gravity from Dirichlet boundary conditions}},  {\em
  Phys. Lett.} {\bf B688} (2010) 137--141,
  [\href{http://xxx.lanl.gov/abs/0910.5474}{{\tt arXiv:0910.5474}}].

\bibitem{deRham:2010ik}
C.~de~Rham and G.~Gabadadze, {\it {Generalization of the Fierz-Pauli Action}},
  {\em Phys. Rev.} {\bf D82} (2010) 044020,
  [\href{http://xxx.lanl.gov/abs/1007.0443}{{\tt arXiv:1007.0443}}].

\bibitem{deRham:2010kj}
C.~de~Rham, G.~Gabadadze, and A.~J. Tolley, {\it {Resummation of Massive
  Gravity}},  {\em Phys. Rev. Lett.} {\bf 106} (2011) 231101,
  [\href{http://xxx.lanl.gov/abs/1011.1232}{{\tt arXiv:1011.1232}}].

\bibitem{deRham:2014zqa}
C.~de~Rham, {\it {Massive Gravity}},  {\em Living Rev. Rel.} {\bf 17} (2014) 7,
  [\href{http://xxx.lanl.gov/abs/1401.4173}{{\tt arXiv:1401.4173}}].

\bibitem{deRham:2016nuf}
C.~de~Rham, J.~T. Deskins, A.~J. Tolley, and S.-Y. Zhou, {\it {Graviton Mass
  Bounds}},  \href{http://xxx.lanl.gov/abs/1606.0846}{{\tt arXiv:1606.0846}}.

\bibitem{Burrage:2011cr}
C.~Burrage, C.~de~Rham, L.~Heisenberg, and A.~J. Tolley, {\it {Chronology
  protection in galileon models and massive gravity}},  {\em JCAP} {\bf 1207}
  (2012) 004, [\href{http://xxx.lanl.gov/abs/1111.5549}{{\tt
  arXiv:1111.5549}}].

\bibitem{Cheung:2016yqr}
C.~Cheung and G.~N. Remmen, {\it {Positive Signs in Massive Gravity}},  {\em
  JHEP} {\bf 04} (2016) 002, [\href{http://xxx.lanl.gov/abs/1601.0406}{{\tt
  arXiv:1601.0406}}].

\bibitem{Bonifacio:2016wcb}
J.~Bonifacio, K.~Hinterbichler, and R.~A. Rosen, {\it {Positivity constraints
  for pseudolinear massive spin-2 and vector Galileons}},  {\em Phys. Rev.}
  {\bf D94} (2016), no.~10 104001,
  [\href{http://xxx.lanl.gov/abs/1607.0608}{{\tt arXiv:1607.0608}}].

\bibitem{Bros:1964iho}
J.~Bros, H.~Epstein, and V.~J. Glaser, {\it {Some rigorous analyticity
  properties of the four-point function in momentum space}},  {\em Nuovo Cim.}
  {\bf 31} (1964) 1265--1302.

\bibitem{Martin:1965jj}
A.~Martin, {\it {Extension of the axiomatic analyticity domain of scattering
  amplitudes by unitarity. 1.}},  {\em Nuovo Cim.} {\bf A42} (1965) 930--953.

\bibitem{Froissart:1961ux}
M.~Froissart, {\it {Asymptotic behavior and subtractions in the Mandelstam
  representation}},  {\em Phys. Rev.} {\bf 123} (1961) 1053--1057.

\bibitem{Martin:1962rt}
A.~Martin, {\it {Unitarity and high-energy behavior of scattering amplitudes}},
   {\em Phys. Rev.} {\bf 129} (1963) 1432--1436.

\bibitem{Jin:1964zza}
Y.~S. Jin and A.~Martin, {\it {Number of Subtractions in Fixed-Transfer
  Dispersion Relations}},  {\em Phys. Rev.} {\bf 135} (1964) B1375--B1377.

\bibitem{Goon:2016ihr}
G.~Goon, K.~Hinterbichler, A.~Joyce, and M.~Trodden, {\it {Aspects of Galileon
  Non-Renormalization}},  {\em JHEP} {\bf 11} (2016) 100,
  [\href{http://xxx.lanl.gov/abs/1606.0229}{{\tt arXiv:1606.0229}}].

\bibitem{deRham:2013hsa}
C.~de~Rham, M.~Fasiello, and A.~J. Tolley, {\it {Galileon duality}},  {\em
  Phys.Lett.} {\bf B733} (2014) 46--51,
  [\href{http://xxx.lanl.gov/abs/1308.2702}{{\tt arXiv:1308.2702}}].

\bibitem{deRham:2014lqa}
C.~De~Rham, L.~Keltner, and A.~J. Tolley, {\it {Generalized galileon duality}},
   {\em Phys. Rev.} {\bf D90} (2014), no.~2 024050,
  [\href{http://xxx.lanl.gov/abs/1403.3690}{{\tt arXiv:1403.3690}}].

\bibitem{Bellazzini:2016xrt}
B.~Bellazzini, {\it {Softness and amplitudes positivity for spinning
  particles}},  {\em JHEP} {\bf 02} (2017) 034,
  [\href{http://xxx.lanl.gov/abs/1605.0611}{{\tt arXiv:1605.0611}}].

\bibitem{deRham:2014wfa}
C.~de~Rham and R.~H. Ribeiro, {\it {Riding on irrelevant operators}},
  \href{http://xxx.lanl.gov/abs/1405.5213}{{\tt arXiv:1405.5213}}.

\bibitem{Dvali:2010jz}
G.~Dvali, G.~F. Giudice, C.~Gomez, and A.~Kehagias, {\it {UV-completion by
  classicalization}},  {\em JHEP} {\bf 1108} (2011) 108,
  [\href{http://xxx.lanl.gov/abs/1010.1415}{{\tt arXiv:1010.1415}}].

\bibitem{Dvali:2010ns}
G.~Dvali and D.~Pirtskhalava, {\it {Dynamics of unitarization by
  classicalization}},  {\em Phys.Lett.} {\bf B699} (2011) 78--86,
  [\href{http://xxx.lanl.gov/abs/1011.0114}{{\tt arXiv:1011.0114}}].

\bibitem{Dvali:2011nj}
G.~Dvali, {\it {Classicalize or not to Classicalize?}},
  \href{http://xxx.lanl.gov/abs/1101.2661}{{\tt arXiv:1101.2661}}.

\bibitem{Dvali:2011th}
G.~Dvali, C.~Gomez, and A.~Kehagias, {\it {Classicalization of Gravitons and
  Goldstones}},  {\em JHEP} {\bf 1111} (2011) 070,
  [\href{http://xxx.lanl.gov/abs/1103.5963}{{\tt arXiv:1103.5963}}].

\bibitem{Vikman:2012bx}
A.~Vikman, {\it {Suppressing Quantum Fluctuations in Classicalization}},  {\em
  Europhys.Lett.} {\bf 101} (2013) 34001,
  [\href{http://xxx.lanl.gov/abs/1208.3647}{{\tt arXiv:1208.3647}}].

\bibitem{Kovner:2012yi}
A.~Kovner and M.~Lublinsky, {\it {Classicalization and Unitarity}},  {\em JHEP}
  {\bf 1211} (2012) 030, [\href{http://xxx.lanl.gov/abs/1207.5037}{{\tt
  arXiv:1207.5037}}].

\bibitem{Keltner:2015xda}
L.~Keltner and A.~J. Tolley, {\it {UV properties of Galileons: Spectral
  Densities}},  \href{http://xxx.lanl.gov/abs/1502.0570}{{\tt
  arXiv:1502.0570}}.

\bibitem{Deffayet:2011gz}
C.~Deffayet, X.~Gao, D.~Steer, and G.~Zahariade, {\it {From k-essence to
  generalised Galileons}},  {\em Phys.Rev.} {\bf D84} (2011) 064039,
  [\href{http://xxx.lanl.gov/abs/1103.3260}{{\tt arXiv:1103.3260}}].

\end{thebibliography}\endgroup
%%%%%%%%%%%%%%%%

\end{document}